\documentclass[graybox]{svmult}
\usepackage{mathptmx}       % selects Times Roman as basic font
\usepackage{helvet}         % selects Helvetica as sans-serif font
\usepackage{courier}        % selects Courier as typewriter font
\usepackage{type1cm}        % activate if the above 3 fonts are not available on your system
\usepackage{cite}
\usepackage{makeidx}         % allows index generation
\usepackage{graphicx}        % standard LaTeX graphics tool when including figure files
\usepackage{multicol}        % used for the two-column index
\usepackage[bottom]{footmisc}% places footnotes at page bottom
\usepackage{enumerate}
\usepackage{hyperref}
\usepackage{url}
\newcommand{\nid}{\noindent}

\newcommand{\beq}{\begin{equation}}
\newcommand{\eeq}{\end{equation}}
\newcommand{\ds}{\displaystyle}

\makeindex             % used for the subject index
\begin{document}

\title*{Free fall and self-force: an historical perspective}
\titlerunning{Free fall} 
\author{Alessandro Spallicci}
\authorrunning{A. Spallicci} 
\institute{{Alessandro D.A.M. Spallicci di Filottrano}
{\at \\ 
{Observatoire des Sciences de l'Univers en R\'egion Centre, Universit\'e d'Orl\'eans}\\
{LPC2E - Campus CNRS, 3A Avenue de la Recherche Scientifique 45071 Orl\'eans France}\\
\email{spallicci@cnrs-orleans.fr}}}

\maketitle

\abstract{Free fall has signed the greatest markings in the history of physics through the leaning Pisa tower, the Cambridge apple tree and the Einstein lift. The perspectives offered by the capture of stars by supermassive black holes are to be cherished, because the study of the motion of falling stars will constitute a giant step forward in the understanding of gravitation in the regime of strong field. After an account on the perception of free fall in ancient times and on the behaviour of a gravitating mass in Newtonian physics, this chapter deals with last century debate on the repulsion for a Schwarzschild black hole and mentions the issue of an infalling particle velocity at the horizon. Further, black hole perturbations and numerical methods are presented, paving the way to the introduction of the self-force and other back-action related methods. 
The impact of the perturbations on the motion of the falling particle is computed via the tail, the back-scattered part of the perturbations, or via 
a radiative Green function. In the former approach, the self-force acts upon the background geodesic; in the latter, the geodesic is conceived 
in the total (background plus perturbations) field. Regularisation techniques (mode-sum and Riemann-Hurwitz $z$ function) intervene to cancel 
divergencies coming from the infinitesimal size of the particle.  
An account is given on the state of the art, including the last results obtained in this most classical problem, together with a perspective encompassing future space gravitational wave interferometry and head-on particle physics experiments. As free fall is patently non-adiabatic, it requires the most sophisticated techniques for studying the evolution of the motion. In this scenario, the potential of the self-consistent approach, by means of which the background geodesic is continuously corrected by the self-force contribution, is examined. }

\section{Introduction}

The two-body problem in general
relativity remains one of the most interesting problems, being still partially unsolved. 
Specifically, the free fall, one of the eldest and classical problems in physics, has characterised the thinking of the most genial developments and it is taken as reference to measure our progress in the knowledge of gravitation.       
Free fall contains some of the most fundamental questions on relativistic motion. The mathematical simplification, given by the reduction to a 2-dimensional case, and the non-likelihood of an astrophysical head-on collision should not throw a shadow on the merits of this problem. Instead, it may be seen as an arena where to explore part of the relevant features that occur to general orbits, e.g. the coupling between radial and time coordinates. 

Although it is easily argued that radiation reaction has a modest impact on radial fall due to the feebleness of cumulative effects (anyhow, in case of high or even 
relativistic - a fraction of $c$ - initial velocity of the falling particle, it is reasonable to suppose a non-modest impact on the waveform and possibly the existence of a signature), it would be presumptuous to consider free fall simpler than circular orbits, or even elliptic orbits if in the latter adiabaticity may be evoked. Adiabaticity has been variously defined in the literature, but on the common ground of the secular effects of radiation reaction occurring on a longer time scale than the orbital period. One definition refers to the particle moving anyhow, although radiating, on the background geodesic (local small deviations approximation), of obviously no-interest herein; another, currently debated for bound orbits, to the secular changes in the orbital motion being stemmed solely by the dissipative effects (radiative approximation); the third to the radiation reaction time scale being much longer than the orbital period (secular approximation), which is a rephrasing of the basic assumption. 

But in radial fall such an orbital period doesn't exist. And as the particle falls in, the problem becomes more and more complex. 
In curved spacetime, at any time the emitted radiation may backscatter off the spacetime curvature, and interact back with the particle later on. Therefore, the instantaneous conservation of energy is not applicable and the momentary self-force acting on the particle depends on the particle's entire history. There is an escape route, though, for periodic motion. But energy-momentum balance can't be evoked in radial fall, lacking the opportunity of any adiabatic averaging. The particle reaction to its radiation has thus to be computed and implemented immediately to determine the effects on the subsequent motion. It is a no-compromise analysis, without shortcuts. Thus, the computation and the application of the back-action all along the trajectory and the continuous correction of the background geodesic, it is the only semi-analytic way to determine motion in non-adiabatic cases. And once this self-consistent approach shall be mastered for radial infall, where simplification occurs for the two-dimensional nature of the problem, it shall be applicable to generic orbits.  

It is worth reminding that the non-adiabatic gravitational waveforms are one of the original aims of the self-force community, since they express i) the physics closer to the black hole horizon; ii) the most complex trajectories; iii) the most tantalising theoretical questions.   

The head-on collisions of black holes and the associated radiation reaction were evoked recently in the context of particle accelerators and thereby showing the richness of the applicability of the radial trajectory also beyond the astrophysical realm. 
As gravity is claimed by some authors to be the dominant force in the transplanckian region, the use of general relativity is adopted for their analysis.

This chapter reviews the problem of free fall of a small mass into a large one, from the beginning of science, whatever this may mean, to the application of the self-force and of a concurring approach, in the last fourteen years. There is no pretension of exhaustiveness and, furthermore, justifiably or not, from this review some topics have been disregarded, namely: any orbit different from radial fall; radiation reaction in electromagnetism; but also the head-on of comparable masses and Kerr geometry, post-Newtonian (pN) and effective one-body (EOB) methods; quantum corrections to motion. 

Herein, the terms of self-force and radiation reaction are used rather loosely, though the latter does not include non-radiative modes. 
Thus, the self-force describes any of the effects upon an object's motion which are proportional to its own mass. Nevertheless, to the term self-force is often associated a specific method and it is preferable to adopt the term back-action whenever such association is not meant.  
  
Geometric units ($G = c = 1$) and the convention (-,+,+,+) are adopted, unless otherwise stated. The full metric is given by ${\bar g}_{\alpha\beta} (t,r) = g_{\alpha\beta}(r) + h_{\alpha\beta} (t,r)$ where $g_{\alpha\beta}$ is the background metric of a black hole of mass $M$ and
$h_{\alpha\beta}$ is the perturbation caused by a test particle of mass $m$. 

\section{The historical heritage}

The analysis of the problem of motion certainly did not start with a refereed publication and it is arduous to identify individual contributions. Therefore, an arbitrary and convenient choice has led to select only renowned names. 

%% Around 600 B.C., Anaximandros (see Aristot\'el\=es \cite{ar}, as only one original fragment of Anaximandros has survived) considered the Earth %% floating very still in the infinite universe, not supported by anything, and remaining in the same place because of its indifference. Two centuries later, 
Aristot\'el\=es in the fourth century B.C. 
%% put again the Earth at the centre of a finite universe. He 
analysed motion qualitatively rather than quantitatively, but he was certainly more geared to a physical language than his predecessors. His views are scattered through his works, though mainly exposed in the Corpus 
Aristotelicum, collection of the works of Aristot\'el\=es, that has survived from antiquity through Medieval manuscript transmission \cite{ar}. He held that there are two kinds of motion for inanimate matter, natural and unnatural. Unnatural motion is when something is being pushed: in this case the speed of motion is proportional to the force of the push. Natural motion is when something is seeking its natural place in the universe, such as a stone falling, or fire rising. For the natural motion of objects falling to Earth, Aristot\'el\=es asserted that the speed of fall is proportional to the weight, and inversely proportional to the density of the medium the body is falling through. He added, though, that there is some acceleration as the body approaches more closely its own element; the body increases its weight and speeds up. The more tenuous a medium is, the faster the motion. If an object is moving in void, Aristot\'el\=es believed that it would be moving infinitely fast.

After two centuries, Hipparkhos said, through Simplikios \cite{si}, that bodies falling from high do experience a restraining factor which accounts for the slower movement at the start of the fall. 

Gravitation was a domain of concern in the flourishing Islamic world between the ninth and the thirteenth century by ibn Sh{\=a}kir, al-B{\=\i}r{\=u}n{\=\i}, al-Haytham, al-Khazini. It is doubtful whether gravitation was in their minds in the form of a mutual attraction of all existing bodies, but the debate acquired
significant depth, although not benefiting of any experimental input. Conversely, in Islamic countries experiments were performed as deemed necessary for the development of science.     
In this sense, there was a large paradigmic shift with respect to Greek philosophers, more oriented to abstract speculations. 

Leonardo da Vinci\footnote{Leonardo spent his final years at Amboise, nowadays part of the French R\'egion Centre, under invitation of Fran\c{c}ois I, King of France and Duke of Orl\'eans.} stated that each object doesn't move by its own, and when it moves, it moves under an unequal weight (for a higher cause); and when the wish of the first engine stops, immediately the second stops \cite{lv}. Further, in the context of 15$^{th}$ century gravitation, da Vinci compared planets to magnets for their mutual attraction.

Perception of the beginning of modern science in the early seventeenth century is connected on one hand to a popular legend, according to which Galilei dropped balls of various densities from the Tower of Pisa, and found that lighter and heavier ones fell at the same speed (in fact, he did quantitative experiments with balls rolling down an inclined plane, a form of falling that is slow enough to be measured without advanced instruments); on the other hand, modern science developed when the natural philosophers abandoned the search for a cause of the motion, in favour of the search for a law describing such motion. 
The law of fall, stating that distances from rest are as the squares of the elapsed times, appeared already in 1604 \cite{ga1604} and further developed in two 
famous essays \cite{ga1632, ga1638}. 
  
After another century, another legend is connected to Newton \cite{wh97} who indeed himself told that he was inspired to formulate his theory of gravitation  \cite{ne1687} by watching the fall of an apple from a tree as reported by W. Stuckeley and J. Conduit. The fatherhood of the inverse square law, though, was claimed by R. Hooke and it can be traced back even further in the history of physics.   

The pre-Galilean physics, see Drake \cite{dr00}\footnote{His book presents the contributions by several less known researchers in the flow of time, being a well argued and historical - but rather uncritical - account. An other limitation is the neglect of non-Western contributions to the development of physics.}, had an insight on phenomena that is not to be dismissed at once. For instance, the widespread belief that fall is unaffected by the mass of the falling object shall be examined throughout the chapter, through the concept of Newtonian back-action and through the general relativistic analysis of  the capture of stars by supermassive black holes. 

\section{Uniqueness of acceleration and the Newtonian back-action}

One of the most mysterious and sacred laws in general relativity is the equivalence principle (EP). Confronted with ``the happiest thought'' of Einstein's life, it is a relief for those who adventure into its questioning to find out that notable relativists share this humble opinion\footnote{ 
Indeed, it has been stated by Synge \cite{sy60} ``...Perhaps they speak of the principle of equivalence. If so, it is my turn to have a blank mind, for I have never been able to understand this principle...''.}.
This principle is variously defined and here below some most popular versions are listed:

\begin{enumerate}[I]

\item{All bodies equally accelerate under inertial or gravitational forces.}
\item{All bodies equally accelerate independently from their internal composition.}

\vskip 5pt

{\nid In general relativity, the language style gets more sophisticated:}  

\vskip 5pt

\item{At every spacetime point of an arbitrary gravitational field, it is possible to choose a locally inertial coordinate system such that the laws of nature take the same form as in an unaccelerated coordinate system. The laws of nature
concerned might be all laws (strong EP), or solely those dealing with inertial motion (weak EP) or all laws but those dealing with inertial motion (semi-strong EP).}
\item{A freely moving particle follows a geodesic of spacetime.}

\end{enumerate}

It is evident that both conceptually and experimentally, the above different statements are not necessarily equivalent\footnote{For a review on experimental status of these fundamental laws, see Will's classical references \cite{wi93, wi06}, or else L\"ammerzahl's alternative view \cite{la08}, while the relation to energy conservation is analysed by Haugan \cite{ha79}}, although they can be connected to each other (for instance: the EP states that the ratio of gravitational mass to inertial mass is identical for all bodies and convenience suggests that this ratio is posed equal to unity). 
In this chaper, only the fourth definition will be dealt with\footnote{For the first definition, it is worth mentioning the following observation \cite{sy60} ``...Does it mean that the effects of a gravitational field are indistinguishable from the effects of an observer's acceleration? If so, it is false. In Einstein's theory, either there is a gravitational field or there is none, according as the Riemann tensor does not or does vanish. This is an absolute property; it has nothing to do with any observer's world-line. Space-time is either flat or curved...''
Patently, the converse is also far reaching: if an inertial acceleration was strictly equivalent to one produced by a gravitational field, curvature would be then associated to inertial accelerations. Rohrlich \cite{ro00} stresses that the gravitational field must be static and homogeneous and thus in absence of tidal forces. But no such a gravitational field exists or even may be conceived! Furthermore, the particle internal structure has to be neglected. 

The second definition is under scrutiny by numerous experimental tests compelled by modern theories as pointed out by Damour \cite{da09a} and Fayet \cite{fa01}. 

First and last two definitions are correct in the limit of a point mass. An interesting discussion is offered by Ciufolini and Wheeler \cite{ciwh95} on the non-applicability of the concept of a locally inertial frame (indeed
a spherical drop of liquid in a gravity field would be deformed by tidal forces after some time as a state of the art gradiometer may reach  sensitivities such to detect the tidal forces of a weak gravitational field in a freely falling cabin). Mathematically, locality, for which the metric tensor $g_{\mu\nu}$ reduces to the Minkowski metric and the first derivatives of the metric tensor are zero, is limited by the non-vanishing of the Riemann curvature tensor, as in general certain combinations of the second derivatives of $g_{\mu\nu}$ cannot be removed. Pragmatically, it may be concluded that violating effects on the EP may be negligible in a sufficiently small spacetime region, close to a given event.} and interestingly, it can 
be reformulated, see Detweiler and Whiting \cite{dewh03, whde03}, in terms of geodesic motion in the perturbed field. Then, the back-action results into the geodesic motion of the particle in the metric 
$ g_{\alpha\beta}(r) + h_{\alpha\beta}^R (t,r)$ where $h_{\alpha\beta}^R$ is the regular part of the perturbation caused by a test particle of mass $m$. 
Thus, the concept of geodesic motion is adapted to include the influence of $m$ through $h_{\alpha\beta}^R$. 

A teasing paradox concerning radiation has been conceived relative to a charge located in an Earth orbiting spacecraft. Circularly moving charges do radiate, but relative to the freely falling space cabin the charge is at rest and thus not radiating. Ehlers \cite{ri06} solves the paradox by proposing that ``It is necessary to restrict the class of experiments covered by the EP to those that are isolated from bodies of fields outside the cabin''. The transfer of this paradox to the
gravitational case, including the case of radial fall, is immediate.

The EP is receptive of another criticism directed at the relation between the foundations of relativity and their 
implementation: it is somehow confined to the introduction of general relativity, while, for the development of the theory, a  student of general relativity may be rather unaware of it\footnote{Again, this opinion is comforted \cite{sy60} ``...the principle of equivalence performed the essential office of midwife at the birth of general relativity...I suggest that the midwife be now buried with appropriate honours...''.}. 

A popular but wrong interpretation of the EP states that all bodies fall with the same acceleration independently from the value of their mass (sometimes referred as the uniqueness of acceleration). This view is portrayed or vaguely referred to in some undergraduate textbooks, and anyhow largely present in various websites.   
Concerning the uniqueness of acceleration, non-radiative relativistic modes in a circular orbit were analysed by Detweiler and Poisson \cite{depo04}, who showed how the low multipole contributions to the gravitational self-acceleration may produce physical effects, within gauge arbitrariness ($l=0$ determines a mass shift, $l=1$ a centre of mass shift). In \cite{depo04}, the stage is set by a discussion on the gravitational self-force in Newtonian theory for a circular orbit.  Herein, exactly the same pedagogical demonstration of theirs is applied to free fall.  

A small particle of mass $m$ is in the gravitational field of a much larger mass $M$. The origin of the coordinate system coincides with the centre of mass.  
The positions of $M$, $m$ and a field point $P$ are given by $\overrightarrow{\rho}$, $\overrightarrow{R}$
and $\overrightarrow{r}$, respectively (the absolute value of $\overrightarrow{r}$ is $r$ and
$ m \overrightarrow{R}+ M\overrightarrow{\rho}= 0 $).  
In case of the sole presence of $M$, the potential and the acceleration at $P$ are given by:
\beq 
\Phi_0(r)=-\frac{M}{r}, 
~~~~~~~~~~~~~~~~~~~~
\overrightarrow{g}_0(r)=-\nabla
\Phi_0(r)=-\frac{M}{r^3}\overrightarrow{r}. 
\eeq
If $m$ is also present, $M$ is displaced from
the origin and the potential is:

\beq
\Phi(r)=-\frac{M}{|\overrightarrow{r}-\overrightarrow{\rho}|}-\frac{m}{|\overrightarrow{r}-\overrightarrow{R}|}.
\label{eq:phir}
\eeq

Since $m \ll M$, eq. (\ref{eq:phir}) is rewritten in the form of a small variation, that is $\Phi(r)=\Phi_0(r)+\delta\Phi(r)$ or else $\delta\Phi(r)=\Phi(r)-\Phi_0(r)$; 
thus:
\beq
\delta\Phi(r)=
\underbrace{-\frac{m}{|\overrightarrow{r}-\overrightarrow{R}|}}_{\Phi_S}~\underbrace{- \frac{M}{|\overrightarrow{r}-\overrightarrow{\rho}|}+ \frac{M}{r}}_{\Phi_R}. 
\label{dphi}
\eeq
The potential $\delta\Phi(r)$ determines a field that exerts a force on $m$, that is the back-action of the particle. The singular term $ \Phi_S $ diverges, but isotropically around the particle position and thus not contributing to the particle motion. Instead, the remaining regular part acts on the particle.
Since $m\ll M$, the regular parts of the potential and of the acceleration, being $\nabla = {\partial _r} (\overrightarrow{r}/{r})$, are:

\beq
%% \[ 
\Phi_R(r) 
%% = 
%% -\frac{M}{|\overrightarrow{r}-\overrightarrow{\rho}|} =
%% -\frac{M}{|\overrightarrow{r}+\ds\frac{m}{M}\overrightarrow{R}|}=
%% -\frac{M}{|\overrightarrow{r}||1+\ds\frac{m}{M}\ds\frac{\overrightarrow{R}
%% \overrightarrow{r}}{r^2}|}\simeq
%% \]
%% \beq
\simeq -\frac{M}{r} \left [1-\frac{m}{M}
\frac{\overrightarrow{R} \cdot \overrightarrow{r}}{r^2} \right ] + \frac{M}{r} =
m \frac{\overrightarrow{R} \cdot \overrightarrow{r}}{r^3},
\label{phir}
\eeq
\beq 
\overrightarrow{g}_R(r)=-\nabla \Phi_R(r) = 
%% \frac{m}{r^3} \nabla (\overrightarrow{R}
%% \overrightarrow{r})+m \overrightarrow{R} \overrightarrow{r} \nabla
%% \left(\frac{1}{r^3}\right) =
m \frac{3 (\overrightarrow{R} \cdot \overrightarrow{r}) \overrightarrow{r}-r^2
\overrightarrow{R}}{r^5}. 
\eeq
At the particle position, the two components of the acceleration are: 
\beq 
\overrightarrow{g}_R(R)=2 m \frac{\overrightarrow{R}}{R^3},
~~~~~~~~~~~~~~~~~~~~~~
\overrightarrow{g}_0(R)=-M\frac{\overrightarrow{R}}{R^3},
\eeq
and finally the total acceleration is given by (in vector and scalar form):
\beq
\overrightarrow{g}(R)=\overrightarrow{g}_0(R)+\overrightarrow{g}_R(R)=-\frac{M
- 2 m}{R^3}\overrightarrow{R}, 
\eeq
\beq
{g}(R)= - \frac{M}{R^2}\left(1 - 2\frac{m}{M}\right). 
\eeq
The Newtonian back-action of a particle of mass $m$ falling into a much larger mass $M$ is expressed as a correction to the classical value.  
This result is more easily derived, if the partition between singular and regular parts and the vectorial notation is left aside. The force exerted on $m$ is (the origin of coordinate
system is made coincident with the centre of mass for simplicity, so that $m R = M\rho$):
\beq
m {\ddot R} = - \frac{M m}{(\rho + R)^2}, 
\label{accnewton}
\eeq
and thus
\beq
{\ddot R} = - \frac{M}{R^2\left(1 + \ds\frac{\rho}{R}\right)^2} 
%% = - \frac{M}{R^2\left(1 + \ds\frac{m}{M}\right)^2} 
\simeq - \frac{M}{R^2}\left(1 - 2 \frac{m}{M}\right),
\eeq
and
%% \beq
%% m {\ddot R} = - \frac{Mm}{(\rho + R)(\rho + R)} 
%% \eeq
%% and thus:
\beq
{\ddot R} = - \frac{M}{R(\rho + R)\left(1 + \ds\frac{\rho}{R}\right)} 
% = \frac{M}{R(\rho + R)\left(1 + \ds\frac{m}{M}\right)}  
\simeq
- \frac{M}{R(\rho + R)}\left(1 - \frac{m}{M}\right). 
\eeq
%% and then 
%% \beq
%% \simeq
%% - \frac{M}{R^2}\left(1 - 2\frac{m}{M}\right)
%% \eeq
It appears from the preceding computations that the falling mass is slowed down by a factor (1 or 2) proportional to its own mass and dependent upon the measurement approach adopted.       
It may be argued that the $m/M$ term arises because the computation is referred to the centre of mass, but to shift the centre of mass to the centre of $M$ is equivalent to deny the influence of $m$. 

But instead, what about the popular belief that heavier objects fall faster? Let us consider the mass $m$ at height $h$ from the soil and the Earth radius $R_\oplus$; since $\rho + R = h + R_\oplus$, it is found 
%% from the centre of mass expression and from eq. (\ref{accnewton}), 
%% it is easy to work out, without approximations, 
that:
\beq
{\ddot h} = - \frac{M}{\left (h + R_\oplus\right )^2} \left(1 + \frac{m}{M}\right).
\label{popbe1}
\eeq  
The mass is now falling faster thanks to a different observer system and the popular belief appears being confirmed. On the other hand, in a coordinate system whose origin is coincident and comoving with the centre of mass of the larger body $M$, any back-action effect disappears. For $d$, the distance between the two bodies, it is well known that:
\beq  
m{\ddot d} = \frac{mM}{d^2}.   
\eeq
Nevertheless, the translational speed of the moving centre of mass of $M$ (if the latter is fixed, any influence of $m$ is automatically ruled out) is depending upon the value of $m$ and the same applies to eq. (\ref{accnewton}): it is not possible to find an universal reference frame in which the centre of mass moves equally for all various falling masses.   
Thus, the uniqueness of acceleration is result of an approximation, although often portrayed as an exact statement\footnote{The difference between fall in vacuum and in the air has been the subject of a polemics between the former French Minister of Higher Education and Research Claude All\`egre and the Physics Nobel Prize Georges Charpak, solicited by the satirical weekly `Le Canard Encha\^in\'e' \cite{ce99}. The Minister affirmed on French television in 1999
``Pick a student, ask him a simple question in physics: take a petanque and a tennis ball, release them; which one arrives first? The student would tell you: ``the petanque''. Hey no, they arrive together; and it is a fundamental problem, for which 2000 years were necessary to understand it. These are the basis that everyone should know.'' The humourists wisecracked that the presence of air would indeed prove the student being right and tested their claim by means of filled and empty plastic water bottles being released from the second floor of their editorial offices ...and asked the Nobel winner to compute the difference due to the air, whose influence was denied by the Minister. But in this polemics, no one drew the attention to the Newtonian back-action, also during the polemics revamped in 2003 by All\`egre \cite{al03} who compared this time a heavy object and a paper ball. Such forgetfulness or misconception is best represented by the Apollo 15 display of the simultaneous fall of a feather and a hammer \cite{apollo15}.}, or else consequence of gauge choice\footnote{During the Bloomington 2009 Capra meeting, this state of affairs was presented as `the confusion gauge'.}. 
The uniqueness of acceleration holds as long as the values of the masses of the falling bodies are negligible. Correctly stated, the principle hardly sounds like a principle: all bodies fall with the same acceleration independently from their mass, if ... we neglect their mass. Although the preceding is elementary, 
misconceptions tend to persist in colleges and higher education.   

It is concluded that the Newtonian back-action manifests itself with different numerical factors possibly carrying opposite sign (from the Pisa tower - 100 pisan arms tall -  for an observer situated at its feet, Newtonian back-action shows roughly as proportional to $1.7 \times 10^{-24}~m/s^2$ for each falling kilogram). This feature corresponds to the gauge freedom in general relativity.  

For the latter, when considering perturbations, the energy radiated through gravitational waves is proportional to $m^2/M$ and thus the energy leaking from the nominal motion. Therefore, the concept of uniqueness of acceleration is further affected, as it will be shown further.

Finally, it is quoted  \cite{dewh03} that with only local measurements, the observer
has no means of distinguishing the perturbations from the background metric. In the next section, it is shown that the concept of locality or non-locality of measurements associated to free fall, even without taking into account radiation reaction, is far from being evident and has fueled a controversy for more than $90$ years. 

\section{The controversy on the repulsion and on the particle velocity at the horizon} 

The concept of light being trapped in a star was presented in 1783 by Michell \cite{mi1784} in front of the Royal Society audience and later by 
Laplace \cite{la1796, la1799}. Preti \cite{pr09} describes the close resemblance between the algebraic formulation of Laplace \cite{la1799}
and the concept of a black hole, term coined in 1967 by Wheeler \cite{wh67}. 
In the last century, the Earth, once the attracting mass of reference, was silently replaced by the black hole. But, as many centuries before and after Newtonian gravity were necessary to formulate motion on the Earth, it should not be a surprise 
that it is taking more than a century to resolve the same Newtonian questions in the more complex Einsteinian general relativity on a black hole.      

The existence and the detectability of gravitational waves, the validity of the 
quadrupole formula are among the notorious debates that have characterised general relativity, as described by Kennefick \cite{ke07}. But closer to the topic of this chapter is the, surprisingly since almost endless, controversy on the radial motion in the unperturbed Schwarzschild or properly Schwarzschild-Droste (henceforth SD) metric \cite{sc16, dr16a, dr16b}\footnote{Rothman \cite{ro02} gives a brief historical account on Droste's independent derivation of the same metric published by Schwarzschild, in the same year 1916. Eisenstaedt \cite{ei82} mentions previous attempts by Droste \cite{dr15} on the basis of the preliminary versions of general relativity by Einstein and Grossmann \cite{eigr13}, later followed by the Einstein's works (general relativity was completed in 1915 and first systematically presented in 1916 \cite{ei16c}) and Hilbert's \cite{hi16}. Antoci \cite{an03} and Liebscher \cite{anli01} emphasise Hilbert's \cite{hi17} and Weyl's \cite{we17} later derivations of solutions for spherically symmetric non-rotating bodies. Incidentally, Ferraris, Francaviglia and Reina point to the contributions of Einstein and Grossmann \cite{eigr14}, Lorentz \cite{lo15} and obviously Hilbert \cite{hi16} to the variational formulation.}, intertwined with the early debates on the apparent singularity at $2M$ and on the belief of the impenetrability of this singularity due to the infinite value of pression\footnote{Earman and Eisenstaedt \cite{eaei99} describe the lack of interest of Einstein for singularities in general relativity. The debate 
at the Coll\`ege de France during Einstein's visit in Paris in 1922 included a witty exchange on pression (the Hadamard  
`disaster'), see 
%% Nordmann \cite{no22} and 
Biezunski \cite{bi91}.} at $9/4~M$.   
  
Most references for analysis of orbital motion, e.g. the first comprehensive analysis by Hagihara \cite{ha31} or the later and popular book by Chandrasekhar \cite{ch83}, don't address this debate, that has invested names of the first rank in the specialised early literature.  

An historically oriented essay by Eisenstaedt \cite{ei87} critically scrutinises the relation that relativists have with free fall\footnote{The translation of the title and of the introduction to section 5 of \cite{ei87} serves best this paragraph ``The impasse (or have the relativists fear of the free fall?) [..] the problem of the free fall of bodies in the frame of [..] the Schwarzschild solution. More than any other, this question gathers the optimal conditions of interest, on the technical and epistemological levels, without inducing nevertheless a focused concern by the experts. Though, is it necessary to emphasise that it is a first class problem to which classical mechanics has always showed great concern ... from Galileo; which more is the reference model expressing technically the paradigm of the lift in free fall dear to Einstein? The matter is such that the case is the most elementary, most natural, an extremely simple problem ...apparently, but which raises extremely delicate questions to which only the less conscious relativists believe to reply with answers [..]. Exactly the type of naive question that best experts prefer to leave in the shadow, in absence of an answer that has to be patently clear to be an answer. Without doubts, it is also the reason for which this question induces a very moderate interest among the relativists ...''}. 
This section does not have any pretension of topical (e.g. photons in free fall are not dealt with) or bibliographical completeness. The questions posed in this debate concern the radial fall of a particle into a SD black hole and may be summarised as:
\begin{itemize}
\item{Is there an effect of repulsion such that masses are bounced back from the black hole? Or more mildly, does the particle speed, although always inward, reaches a maximal value and then slows down? And if so, at which speed or at which coordinate radius?}
\item{Does the particle reaches the speed of light at the horizon?}
\end{itemize}

The discussion is largely a reflection of coordinate arbitrariness (and unawareness of its consequences), but the debaters showed sometimes a passionate affection to a coordinate frame they considered more suitable for a `real physical' measurement than other gauges. Further, ill-defined initial conditions at infinity, inaccurate wording (approaching rather than equalling the speed of light), sometimes tortuous reasonings despite the great mathematical simplicity, scarce propension to bibliographic research with consequent claim of historical 
findings \cite{loma09}, they all contributed to the duration of this debate. 
The approach of this section is to cut through any tortuous reasoning \cite{pd08} and show the essence of the debate by means of a clean and simple presentation, thereby paying the price of oversimplification.  
	
Four types of measurements can be envisaged: local measurement of time $dT$, non-local measurement of time $dt$, local measurement of length $dR$, non-local measurement of length $dr$. Locality is somewhat a loose definition, but it hints at those measurements by rules and clocks affected by gravity (of the SD black hole) and noted by capital letters $T,R$, while non-locality hints at measurements by rules and clocks not affected by gravity (of the SD black hole) and noted by small letters $t,r$\footnote{This definition is 
not faultless (there is no shield to gravity), but it is the most suitable to describe the debate, following Cavalieri and Spinelli \cite{casp73, casp77, sp89} and Thirring \cite{th61}.}.   
Therefore, for determining (velocities and) accelerations, four possible combinations do exist: 
\begin{itemize}
\item{Unrenormalised acceleration $d^2r/dt^2$;} 
\item{Semi-renormalised acceleration $d^2R/dt^2$;} 
\item{Renormalised acceleration $d^2R/dT^2$;} 
\item{Semi-renormalised acceleration $d^2r/dT^2$.}
\end{itemize}
The latter hasn't been proposed in the literature and discussion will be limited to the first three types. The former two 
present repulsion at different conditions, while the third one never presents repulsion. 
 
The first to introduce the idea of gravitational repulsion was Droste \cite{dr16a, dr16b} himself. He defines: 

\beq
dR = \frac{dr}{\sqrt{1-\ds\frac{2M}{r}}},
\label{deltadroste}
\eeq
which, after integration, Droste called the distance $\delta$ from the horizon. This quantity is derived from the SD metric posing $dt = 0$, delicate operation since the relation between proper and coordinate times  
varies in space as explained by Landau and Lifshits \cite{lali41}; thus it may be accepted only for a static observer (obviously the notion of static observer raises in itself a series of questions, see e.g. Doughty \cite{do81}, Taylor and Wheeler \cite{tawh00}.). Through a Lagrangian and the relation of eq. (\ref{deltadroste}), for radial trajectories Droste derives that the semi-renormalised velocity and acceleration are given by ($A$ is a constant of motion, equal to unity for a particle falling with zero velocity at infinity):

\beq
\frac{dR}{dt} = - \sqrt{\left(1 - \frac{2M}{r}\right ) \left( 1 - A + \frac{2AM}{r}\right)},
\label{eq:dRdtdroste}
\eeq

\beq
\frac{d^2 R}{dt^2} = - \frac{M}{r^2} 
\left[
\sqrt{1- \frac {2M}{r}} - 
{\ds 
\frac 
{2\left(\ds dR/dt\right)^2}{\sqrt{1- \ds \frac {2M}{r}}}} 
\right]
= \frac{M}{r^2} \left( 1 - 2A + \frac{4AM}{r} \right)\sqrt{1- \frac {2M}{r}}, 
\label{eq:d2Rdt2droste}
\eeq
where the constant of motion $A$ is given by:
\[
A = \left( 1 - \frac{2M}{r}\right)^{-1} - ({dr}/{dt})^2\left( 1 - \frac{2M}{r}\right)^{-3}.
\]

From eq. (\ref{eq:d2Rdt2droste}), two conditions may be derived for the semi-renormalised acceleration, for either of which the repulsion (the acceleration is positive) occurs for $A=1$ if 
$ r<4M $ or else $dR/dt > \sqrt{1/2}\sqrt{1- 2M/r}$. 

Instead in his thesis \cite{dr16a}, Droste investigated the unrenormalised velocity acceleration and for zero velocity at infinity, they are:

\beq
\frac{d r}{dt} = 
- \left(1- \frac {2M}{r} \right )\sqrt{\frac {2M}{r}}, 
\eeq

\beq
\frac{d^2 r}{dt^2} = 
- \frac{M}{r^2} 
\left[
1- \frac {2M}{r} - 
{\ds 
\frac 
{3\left(\ds dr/dt \right)^2}{1- \ds \frac {2M}{r}}} 
\right]=
- \frac{M}{r^2} 
\left(
1- \frac {2M}{r}\right) 
\left( 1- \frac {6M}{r}\right), 
\eeq
for which repulsion occurs if, still for a particle falling from infinity with zero initial velocity, $ r<6M $ or else $dr/dt > 1/\sqrt{3}(1 - 2M/r )$. 

The impact of the choice of coordinates on generating repulsion was not well perceived in the early days of general relativity. Further, many notable authors as Hilbert \cite{hi17, hi24}, Page \cite{pa20}, Eddington \cite{ed20}, von Laue \cite {vl21} in the German original version of his book, Bauer \cite{ba22}, de Jans \cite{dj23, dj24a, dj24b} although indirectly by referring to the German version of \cite{vl21}, arrive independently and largely ignoring the existence of Droste's work, to the same conclusions in semi-renormalised or unrenormalised coordinates. 

The initial conditions\footnote{Generally, the setting of the proper initial conditions may be a delicate issue e.g. when associated with an initial radiation content expressing the previous history of the motion as it will be later discussed; or, in absence of radiation, when an external (sort of third body) mechanism prompting the motion to the two body system is to be taken into account. The latter case is represented by the thought experiment conceived by Copperstock \cite{co74} 
aiming to criticise the quadrupole formula. The experiment consisted in two fluid balls assumed to be in static equilibrium and held apart by a strut, with membranes to contain the fluid, until time $t = 0$.  
Between $t = 0 $ and $t = t_{1}$, the strut and the membranes are dissolved 
and afterwards the balls fall freely. Due to the static initial conditions, 
there is a clear absence of incident radiation, but the behaviour of the fluid
balls in the free fall phase depends on how the transition from the 
equilibrium to the free fall takes place. This initial dependence obscured the debate on the quadrupole formula.}
may astray the particle from being attracted by the gravitating mass. Indeed, Droste \cite{dr16b} and Page \cite{pa20} refer to particles having velocities at infinity equal or larger of $1/\sqrt 2$ for the semi-renormalised coordinates and 
equal or larger of $1/\sqrt 3$ for the unrenormalised coordinates. These conditions dictate to Droste and Page that the particle is constantly slowed down when approaching the black hole and therefore impose to gravitation an endless repulsive action. 

In the later French editions of his book, von Laue \cite{vl21} writes the radial geodesic in proper time, but it is only in 1936 that Drumaux    
\cite{dr36} fully exploits it. Drumaux criticises the use of the semi-renormalised velocity and 
considers eq. (\ref{deltadroste}) as defining the physical measurement of length $dR$. Similarly, the relation between coordinate and proper times (for $dr = 0$) provides the physical measurement of time $dT$:

\beq
dT = \sqrt{1-\frac{2M}{r}} dt. 
\label{drdTdt}
\eeq
Thereby, Drumaux derives the renormalised velocity and acceleration in proper time:

\beq
\frac{d R}{dT} = \sqrt{\frac {2M}{r}},
\eeq

\beq
\frac{d^2R}{dT^2} = - \frac{M}{r^2}\sqrt{1 - \frac {2M}{r}}, 
\eeq
for which no repulsion occurs. This approach is followed by von Rabe \cite{vr47}, Whittaker \cite{wh53}, Srnivasa Rao \cite{sr66}, Zel'dovich and Novikov \cite{zeno67}. Nevertheless, McVittie, almost thirty years after Drumaux \cite{mv56}, still reaffirms that the particle is pushed away by the central body as do Treder \cite{tr72}, also in cooperation with Fritze \cite{trfr75}, Markley \cite{ma73}, Arifov \cite{ar80, ar81}, McGruder \cite{mg82}. 
A discussion on radar and Doppler measurements with semi-renormalised measurements was offered by Jaffe and Shapiro \cite{jash72, jash73}.   
The controversy seems to be extinguished in the 80s, although recent research papers still refer to it, e.g. Kutschera and Zajiczek \cite{kuza09}.
 
For the particle's velocity at the horizon, another, though related, debate has taken place in some of the above mentioned references as well as in Landau and Lifshits \cite{lali41}, Baierlein \cite{ba73}, Janis \cite{ja73, ja77}, Rindler \cite{ri79}, 
Shapiro and Teukolsky \cite{shte83}, Frolov and Novikov \cite{frno98}, Mitra \cite{mi00}, Crawford and Tereno \cite{crte02}, 
M\"{u}ller \cite{mu08} the last ones being recently published. Whether the velocity is $c$ or less, it is still the question posed by these papers. 
   
The further step forward in the analysis of a freely falling mass into a SD black hole has taken place in the period from 1957 to 1997. In these forty years\footnote{Free fall has also been studied in other contexts. Synge \cite{sy60} undertakes a detailed investigation of the problem and shows that, actually, the
gravitational field (i.e. the Riemann tensor) plays an extremely small role in the phenomenon of free fall 
and the acceleration of $980~cm/sec$ is, in fact, due to the curvature of the world line of the tree branch.  The apple is accelerated 
until the stem breaks, then the world line of the 
apple becomes inertial until the ground collides with it.}, the falling mass finally radiates energy (the radiated gravitational power is proportional to the square of the third time derivative of the quadrupole moment which is different than zero), but its motion is still unaffected by the radiation emitted. The influence of the radiation on the motion of a particle of infinitesimal size wasn't dealt with until 1997.  

\section{Black hole perturbations}

Perturbations were first dealt with by Regge and Wheeler \cite{re57, rewh57}, where a SD black hole was shown to regain stability after undergoing small vibrations about its spherical form, if subjected to a small perturbation\footnote{For a critical assessment of black hole stability, see
Dafermos and Rodnianski \cite{daro08}.}. The analysis was carried out thanks to the 
first application to a black hole of the Einstein equation at higher order. 

The SD metric describes the background field $g_{\mu\nu}$ on which the perturbations $h_{\mu\nu}$ arise. It is given by:  

\beq
ds^2 =
-\left(1-\frac{2M}{r}\right)dt^2
+\left(1-\frac{2M}{r}\right)^{-1} dr^2
+r^2\left(d\theta^2+\sin^2\theta d\phi^2\right). 
\label{eq:sdmetric}
\eeq
Eq. (\ref{eq:sdmetric}) originates from the 
Einstein field equation in vacuum, consisting in the vanishing of the Ricci tensor $R_{\mu\nu} = 0$. 

Instead, the Regge-Wheeler equation derives from the vacuum condition, but this time posed on the 
first order variation of the Ricci tensor $\delta R_{\mu\nu} = 0$. 
The generic form of the variation of the Ricci tensor was found by Eisenhart \cite{ei26}
and it is given by: $\delta R_{\mu\nu} = - \delta\Gamma^\beta_{\mu\nu ;\beta} + \delta\Gamma^\beta_{\mu\beta ;\nu}$
where the tensor $\delta\Gamma^\alpha_{\beta\gamma}$, variation of the 
Christoffel symbol (a pseudo-tensor), is:
$\delta\Gamma^\alpha_{\beta\gamma} = 1/2~g^{\alpha\nu}
(h_{\beta\nu;\gamma} + h_{\gamma\nu;\beta} - h_{\beta\gamma;\nu})$, being the perturbation $h_{\mu\nu} = \delta g_{\mu\nu}$.  Replacing the latter in the vanishing variation of the Ricci tensor, a system of ten second order differential equations in 
$h_{\mu\nu}$ was obtained. Exploiting spherical symmetry, finally Regge and Wheeler got a vacuum wave equation out of the three odd-parity equations giving birth to a field that has grown immensely from the end of the 50s\footnote{A well organised introduction, largely based on works by Friedman \cite{fr73} and Chandrasekhar \cite{ch75}, is presented in the already mentioned book by the latter \cite{ch83}. Some selected publications geared to the finalities of this chapter are to be listed: earlier works by Mathews \cite{ma62}, Stachel \cite{st68}, Vishveshvara \cite{vi70}; the relation between odd and parity perturbations \cite{chde75}; the search for a gauge invariant formalism 
by Martel and Poisson \cite{mapo05} complements 
a recent review on gauge invariant non-spherical metric perturbations of the SD black hole spacetimes by Nagar and Rezzolla \cite{nare05}; a classic reference on multiple expansion of gravitational radiation by Thorne \cite{th80}; the derivation by computer algebra by Cruciani \cite{cr00,cr05} of the wave equation governing black hole perturbations; the numerical hyperboloidal approach by Zenginon\u{g}lu \cite{ze10}.}.  

Zel'dovich and Novikov \cite{zeno64} first considered  the problem of 
gravitational waves emitted by bodies moving in the field of a star, on the basis of the  
quadrupole formula, thus at large distances from the horizon, where only a minimal part of the radiation is emitted. 

Whilst a less known semi-relativistic work by Ruffini and Wheeler \cite{ruwh71a,ruwh71b} appeared in the transition from the 60s to the 70s, it was the work by Zerilli \cite{ze70a, ze70b, ze70c}, where the source of perturbations was considered in the form of a radially falling particle, that opened the way to study free fall in a fully, although linearised, relativistic regime at first order. 
The Zerilli equation rules even-parity waves in the presence of a source, i.e. a freely falling point particle,
generating a perturbation for which the difference from the 
SD geometry is small. The energy-momentum 
tensor $T _{\mu\nu}$ is given 
by the integral of the world-line of the particle, the integrand containing  
a four-dimensional invariant $\delta$ Dirac distribution for representation of the point particle trajectory. 
The vanishing of the covariant divergence of $T _{\mu\nu}$ is guaranteed by 
the world-line being a 
geodesic in the background SD geometry; in this way, the problem of 
the linearised theory on flat spacetime (for which the particle moves on a geodesic of flat 
space which determines uniform motion and thereby without emission of radiation) is 
avoided. Finally, the complete description of 
the gravitational waves emitted is given by the symmetric tensor $h_{\mu\nu}$, 
function of $r$, $\theta$, $\phi$ and $t$. 

The formalism can be summarised as follows \cite{ze70a, ze70b, ze70c}\footnote{Two warnings: the literature on perturbations and numerical methods is rather plagued by editorial errors (likely herein too...) and different terminologies for the same families of perturbations. Even parity waves have been named also polar or electric or magnetic, generating some confusion (see the correlation table,  tab. II, in \cite{ze70c}).}. Due to the spherical symmetry of the SD field, the 
linearised field equations for the perturbation $h_{\mu\nu}$ are in the form of rotationally  
invariant operator on $h_{\mu\nu}$, set equal to the energy-momentum tensor 
also expressed in spherical tensorial harmonics: 

\beq
Q[h_{\mu\nu}] \propto T_{\mu\nu}[\delta (z_u)],
\label{qt}
\eeq
where the $\delta(z_u)$ Dirac distribution represents the point particle on the unperturbed trajectory $z_u$.  

The rotational invariance is used to separate out the angular variables in the 
field equations. 
% The solution of such differential equation, based on variables separation, is in the form of a 
% linear combination of the various components $h_{\mu\nu}$:
% \beq
% \sum_{lm} \frac{\Psi_{lm}(t,r)}{r}F[Y_{lm}(\theta,\phi)] 
% \label{eq:psitr}
% \eeq 
% The 
% $F$ operator of eq. (\ref{eq:psitr}) 
% $Q$ operator of eq. (\ref{qt})
% assumes different forms for the 
% scalar, vector and tensor components.
For the spherical symmetry on the 2-dimensional manifold on which $t,r$ are constants under rotation in the $\theta,\phi$ sphere, the ten components of the perturbing symmetric tensor transform like three scalars, two vectors and one tensor:
 
\[
h_{tt}, h_{tr}, h_{rr}
~~~~~~~~
~~~~~~~~
(h_{t\theta}; h_{t\phi}), (h_{r\theta}; h_{r\phi})
~~~~~~~~
~~~~~~~~
\left(\matrix{
h_{\theta\theta} & h_{\theta\phi} \cr 
h_{\phi\theta} & h_{\phi\phi}
}\right ).
\]

In the Regge-Wheeler-Zerilli formalism, the even perturbations (the source term for the odd perturbations vanishes for the radial trajectory and given the rotational invariance through the azimuthal angle, only the index referring to the polar or  latitude angle survives), going as $(-1)^l$, 
are expressed by the following matrix:

{\small
\beq
h_{\mu\nu}\!\! = \!\!\!
\left(\!\!
\matrix{
\left (1-\frac{\displaystyle 2M}{\displaystyle r }\right )H_{0}Y  
& \!\!\!\!\!\!\!\!\!\!\!\!\!\! H_{1}Y 
& \!\!\!\!\!\!\!\!\!\!\! h_0 Y_{,\theta} 
&\!\!\!\!\!\!\!\!\!\!\!\!\! h_0 Y_{,\phi}\cr
sym 
& \!\!\!\!\!\!\!\!\!\!\!\!\! \left (1-\frac{\displaystyle 2M}{\displaystyle r}\right )^{-1}H_{2}Y 
& \!\!\!\!\!\! h_1 Y_{,\theta}   
&\!\!\!\!\!\!\!\!\!\!\! h_1 Y_{,\phi} 
\cr                        
sym
& \!\!\!\!\!\!\!\!\!\!\!\!\!\! sym
& \!\!\!\!\!\!\!\!\!\!\! 
r^2 \left [ K Y + G Y_{,\theta\theta}%\frac{\displaystyle \partial ^ 2}{\displaystyle \partial \theta  ^2} 
\right ] 
&\!\!\!\!\!\!\!\!\!\!\! r^2 G
\left (Y_{,\theta\phi} 
- \cot\theta Y_{,\phi} 
\right ) 
\cr
sym 
& \!\!\!\!\!\!\!\!\!\!\!\!\!\! sym 
& \!\!\!\!\!\!\!\!\!\!\! sym 
& \!\!\!\!\!\!\!\!\!\!\! r^2\sin^2\!\theta\!\!\left[ 
K\!+\!G\!\left(\!\frac{\ds Y_{,\phi\phi}}{\ds\sin^2\!\theta}%\frac{\displaystyle \partial ^ 2}{\displaystyle \partial^2 \phi } 
\!+\! \cot \theta Y_{,\theta}%\frac{\displaystyle \partial} {\displaystyle \partial \theta} \!
\right ) \right ] 
}
\!\!\right ), 
% Y(\theta,\phi ) 
\label{eq:rweven}
\eeq
}
where $H_0, H_1, H_2, h_0, h_1, K, G$ are functions of $(t, r)$ and the $l$ multipole index isn't 
displayed.  
After angular dependence separation, the seven functions of $(t, r)$  
are reduced to four due to a gauge transformation for which $G = h_0 = h_1 = 0$, i.e. the Regge-Wheeler gauge. 

For a point particle of proper mass $m$, represented by a Dirac delta distribution, the
stress-energy tensor is given by:

\beq
T^{\alpha\beta}=
m  \frac{u^\alpha u^\beta}{u^0r^2}~\delta[r - z_u(t)]\delta^2[\Omega], 
\label{emt}
\eeq
where $z_u(t)$ is the trajectory in coordinate time and $u^\alpha$ is the 4-velocity.

Any symmetric covariant tensor can be expanded in spherical harmonics \cite{ze70b}. For radial fall it has been shown that only three  
even source terms do not vanish and that two functions of $(t, r)$ become identical. Finally, six equations are left with three unknown functions $H_0 = H_2, K, H_1$. After considerable manipulation, the following wave equation is obtained (for radial fall $m=0$ and only the index $l$ survives):

\beq
\frac {d^{2} \Psi_l(t,r)}{dr^{*2}} - \frac {d^{2} \Psi_l(t,r)}{dt^2} -
V_{l}(r)\Psi_l (t,r) = S_{l}(t,r),  
\label{eq:rwz*}
\eeq
where $r* = r + 2M\ln (r/2M - 1)$ is the tortoise coordinate; the potential $V_{l}(r)$ is given by: 

\[
V_{l}(r) = \left ( 1- \frac{2M}{r}\right ) \frac {2\lambda^{2}(\lambda +
1)r^{3} + 6\lambda^{2}Mr^{2} + 18\lambda M^{2}r + 18M^{3}} {r^{3}(\lambda r
+ 3M)^2},
\]
being $ \lambda = 1/2(l - 1)(l + 2) $. The source $S_{l}(t,r)$ includes the derivative of the Dirac distribution 
(denoted $\delta '$), coming 
from the combination of the $h_{\mu\nu}$ and their derivatives\footnote{There is an editorial error, a numerical coefficient, in the corresponding expressions (2.16) in \cite{lopr97a} and (2.8) in \cite{lopr97b}, which the footnote 1 at page 3 in \cite{mapo02} doesn't address.}:

\[
S_{l} = \frac{2( r- 2M ) 
\kappa }{r^2(\lambda +1)(\lambda r+3M)}
\times 
\]
\beq
\left \{
\frac{r (r-2M )}{2u^0}
\delta^{\prime }[r-z_u(t)]
- \left [ \frac{r(\lambda + 1)- 3M}{2u^0} - \frac{3Mu^0(r-2M)^2}{r(\lambda r+3M )} \right ]
\delta [r-z_u(t)] 
\right \},
\eeq
for $u^0 = 1/(1 - 2M/z_u)$ being the time component of the 4-velocity and $\kappa = 4m\sqrt{(2l+1)\pi}$. The geodesic in the unperturbed SD metric $z_u(t)$ assumes different forms according to the 
initial conditions\footnote{For a starting point different from infinity or a non-null starting velocity, but not their combination, see Lousto and Price \cite{lopr97a, lopr97b, lopr98}, Martel and Poisson \cite{mapo02}.}; herein, the simplest form is given, namely zero velocity at infinity. Then, $z_u(t)$ is the - numerical - inverse function of: 

\beq
t=-4M\left( {\frac{z_u}{2M}}\right) ^{1/2}-\frac{4M}{3}\left( \frac{z_u}{2M}
\right) ^{3/2}-2M\ln \left[ \frac{\sqrt{\displaystyle{\frac{z_u}{2M}}}-1}
{\sqrt{\displaystyle{\frac{z_u}{2M}}}
+1}\right].   
\label{eq:tofr}
\eeq
The coordinate velocity ${\dot z}_u$ of the particle may be given in terms of its position $z_{u}$: 

\beq
{\dot z}_u = - \left( 1 - \frac{2M}{z_u}\right)
\left( \frac{2M}{z_u} \right)^{1/2}.
\eeq

The dimension of the wavefunction $\Psi$ is such that the energy is proportional to $\int_0^\infty
  \dot\Psi^2\,dt\ $. The wavefunction, in the Moncrief form \cite{mo74} for its gauge invariance, is related to the perturbations via: 

\beq
\Psi_l (t,r)= \frac {r}{\lambda +1}
\left[ K^l+\frac{r-2M}{\lambda r+3M}\left(
H_2^l-r\frac{\partial K^l}{\partial r}\right) \right],
\label{psidef}
\eeq
where the Zerilli \cite{ze70a} normalisation is used for $\Psi_l $. For computations, this allows the choice of a convenient gauge, like the Regge-Wheeler gauge. The inverse relations for the perturbation functions $K$, $H_2$, $H_1$ are given by Lousto \cite{lo00,lona09}:

\beq
K=\frac{6M^2+3M\lambda r+\lambda (\lambda +1)r^2}
{r^2(\lambda r+3M)}\Psi
+\left( 1-\frac{2M}r\right) \,\Psi_{,r}
-\frac{%
\kappa \ u^0(r-2M)^2}{(\lambda +1)(\lambda r+3M)r}\delta 
\label{eq:K},
\eeq

\[ 
H_2=-\frac{9M^3+9\lambda M^2r+
3\lambda ^2Mr^2+\lambda ^2(\lambda +1)r^3}{%
r^2(\lambda r+3M)^2}\,\Psi   
+
\frac{3M^2-\lambda Mr+\lambda r^2}{r(\lambda r+ 3M)}\Psi_{,r} 
+ 
\]
\[
(r-2M)\Psi_{,rr} 
+
\]
\beq
\frac{\kappa u^0(r-2M)[\lambda ^2r^2+2\lambda Mr-3Mr+3M^2]}{r (\lambda +1)(\lambda r+3M)^2}\delta -
\frac{\kappa u^0(r-2M)^2}{
(\lambda +1)(\lambda r+3M)}\delta',  
\label{eq:H02}
\eeq

\beq
H_1 =r \Psi_{,tr}+\frac{\lambda r^2-3M\lambda
r-3M^2}{%
\left( r-2M\right) (\lambda r+3M)}{\Psi_t}- 
\frac{\kappa \ u^0\stackrel{.}{z}_u(\lambda r+M)}{(\lambda +1)(\lambda
r+3M)}\delta 
+
\frac{\kappa \ u^0\stackrel{.}{z}_u r(r-2M)}{(\lambda
+1)(\lambda r+3M)}\delta '. 
\label{eq:H1}
\eeq

Several works by Davies, Press, Price, Ruffini and Tiomno \cite{daruprpr71, daru72, daruti72, ru73a, ru73b, ru78}, but also by individual 
scholars like Chung \cite{ch73}, Dymnikova \cite{dy80} and the forerunners of the Japanese school as Tashiro and Ezawa \cite{taez81}, Nakamura with Oohara and Koijma \cite{naooko87} or with Shibata \cite{shna92}, 
appeared in the frequency domain in the 70s and fewer later on, analysing especially the amplitude and the spectrum of the radiation emitted. Haugan, Petrich, Shapiro and Wasserman \cite{hashwa82, shwa82, peshwa85} modeled the source as 
finite size star of dust.
  
For an infalling mass from infinity at zero velocity, the energy radiated to infinity for all modes \cite{daruprpr71} and the energy absorbed by the black hole \cite{daruti72} for each single mode, and for all modes\footnote{The divergence in summing over all $l$ modes is said to be taken away by considering a finite size particle \cite{daruti72}.} are given by respectively (beware, in 
physical units):

\beq
\sum_l E^r_l = 0.0104 \frac{m^2c^2}{M},
~~~~~~~~~~~~~~~~~~~~~
E^a_l = 0.25 \frac{m^2c^2}{M},
~~~~~~~~~~~~~~~~~~~~~
\sum_l E^a_l = \frac{\pi}{8}mc^2,
\eeq
while most of the energy is emitted below the frequency:

\beq
f_m = 0.08 \frac {c^3}{GM}.
\eeq
Up to $94 \%$ of the energy is radiated between $8M$ and $2M$ and $90 \%$ of it in the quadrupole mode.   
   
Unfortunately, the analysis in the frequency domain doesn't contribute much to the understanding of the particle motion, the limitation having origin in the absence of exact solutions. A Fourier anti-transform of an approximate solution, for instance valid at high frequencies, does not reveal which effect on the motion has the neglect of lower frequencies. Thus, the lack 
of availability of any time domain solution has impeded progress in the comprehension of motion in the perturbative two-body problem. Although studies on analytic solutions were attempted throughout the years, e.g. Fackerell \cite{fa71}, Zhdanov \cite{zh79}, Leaver \cite{le85, le86a, le86b}, Mano, Suzuki and  Takasugi \cite{masuta96} and Fiziev \cite{fi06}, they were limited to the homogeneous equation.   

\section{Numerical solution}

The breakthrough arrived thanks to a specifically tailored finite differences method. It consists of the numerical integration of the inhomogeneous 
wave equation in time domain, proposed by Lousto and Price \cite{lopr97b, lopr98} and based on the mathematical formalism of the particle limit approximation developed in \cite{lopr97a} in the 
Eddington-Finkelstein coordinates \cite{ed24, fi58}. A parametric analysis of the initial data by Martel and Poisson has later appeared \cite{mapo02}. Confirmation of the results, among which the waveforms at infinity, is contained in \cite{ao08}. 

\begin{figure}
\begin{center}
%  Requires \usepackage{graphicx}
\includegraphics[width=8cm]{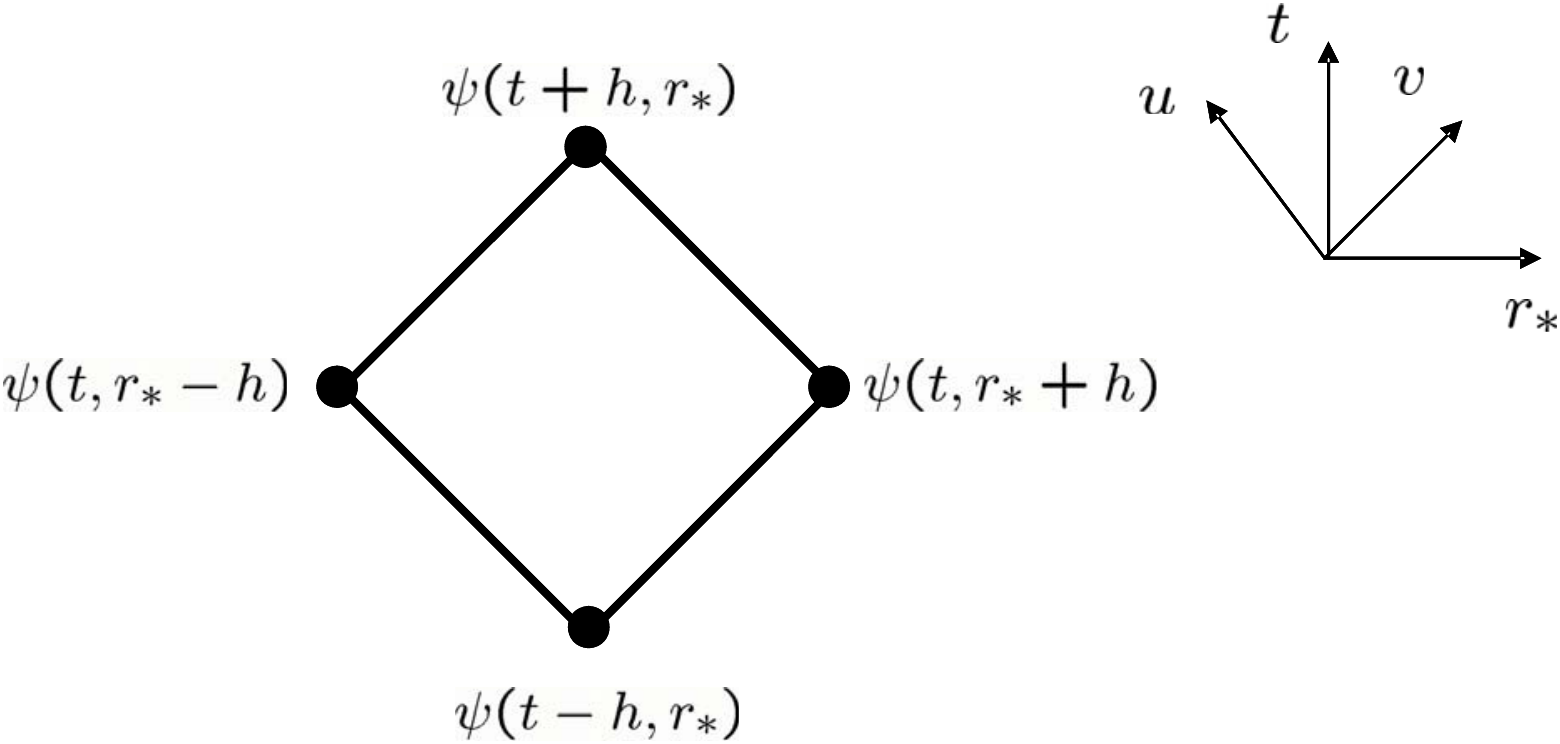}\\
  \caption{An empty cell never crossed by the particle worldline.}\label{cellvide}
\end{center}
\end{figure}

The grid cells are separated in two categories, according to whether the cell is crossed or not by the particle. 
The latter category, fig. \ref{cellvide}, is then formed by the cells for which $r \neq z_u(t)$ and the $\Psi(t,r)$ evolution is not affected by the source. 
It is then sufficient to integrate each term of the homogeneous wave equation. The wave operator allows an exact integration: 

\[
\int\!\!\!\!\int_{\tiny Cell} \left( \partial_{r^*}^2 - \partial_t^2 \right)
\Psi dA = 
\]
\beq
- 4 [ \Psi(t+h, r^*)\\
+\Psi(t-h, r^*) -\Psi(t, r^*+h)-\Psi(t, r^*-h)].
\label{eq:d2psir-d2psit}
\eeq
Instead, the product potential-wavefunction is given by:

\[
\int\!\!\!\!\int_{\tiny Cell} V(r) \Psi dA = 
\]
\beq
V(r)h^2[
\Psi(t+h, r^*) 
+\Psi(t-h, r^*) + \Psi(t, r^*+h) + \Psi(t, r^*-h)+ {\cal O}(h^3) ].
\eeq

The evolution algorithm defines $\Psi$ at the upper cell corner as computed out of the three preceding values:
 \beq
\Psi(t+h,r^*) = - \Psi(t - h,r^*) + 
\left [\Psi(t, r^*+h) + \Psi(t, r^*-h) \right ]\left[1-\frac{h^2}{2}V(r)\right].
\label{schemcelvid}
\eeq

For the cells crossed by the particle, a different integration scheme is imposed, fig. \ref{diffcas}. The product potential-wavefunction is given by: 

\[
\int\!\!\!\!\int_{\tiny Cell} V(r) \Psi dA = V(r) \times 
\]
\beq
\left[A_3 \Psi(t+h, r^*) + A_2\Psi(t-h, r^*) + A_4\Psi(t, r^*+h) + A_1\Psi(t, r^*-h)
+ {\cal O}(h^3)\right],
\eeq
where $A_1, A_2, A_3, A_4$ are the sub-surfaces of the cell. The integration of the source term is given by\footnote{There are editorial errors in the corresponding expressions (3.6) in \cite{lopr97b} and (3.4) in \cite{mapo02}.}: 

\[
\int\!\!\!\!\int_{\tiny Cell} S dA =  
- \int_{t_i}^{t_o}  
\frac{2\kappa (r-2M)}{E (2\lambda + 1)(\lambda z_u + 3M)^2}\times
\]
\[ 
\left[
\frac{6M}{z_u}(1-E^2)+\lambda(\lambda+1)-\frac{3M^2}{z_u^2}+\frac{4\lambda M}{z_u}
\right] dt +
\frac{2\kappa (r-2M)}{E (2\lambda + 1) (\lambda z_u + 3M)}\times
\]
\beq
\left\{{Sign_i} 
\left[ 
1 +\frac{{Sign_i}} {E}\sqrt{\frac{2M}{z_u}-\frac{2M}{z_{u0}}}
\right]^{-1} 
+ { Sign_o} 
\left[
1-\frac{{Sign_o}}{E}\sqrt{
\frac{2M}{z_u}-\frac{2M}{z_{u0}}
}
\right]^{-1} 
\right\},
\eeq
where $t_i$ corresponds to the time of entry of the particle in the cell and $t_o$ the time of departure from the cell; $z_{u0}$ is the initial position of the particle; $E = \sqrt{1 - 2M/z_{u0}}$; ${Sign_i} = + 1 $ if the particle enters the cell on the right, $-1$ if on the left;
${Sign_o} = + 1 $ if the particle leaves the cell on the right, $-1$ if on the left,
fig. \ref{diffcas}.
Through the evolution algorithm, the value of $\Psi$ at the upper cell corner is given by\footnote{There are editorial errors in the corresponding expressions (3.9) in \cite{lopr97b} and (3.5) in \cite{mapo02}.}:

\begin{figure}
\begin{center}
\includegraphics[width=6cm]{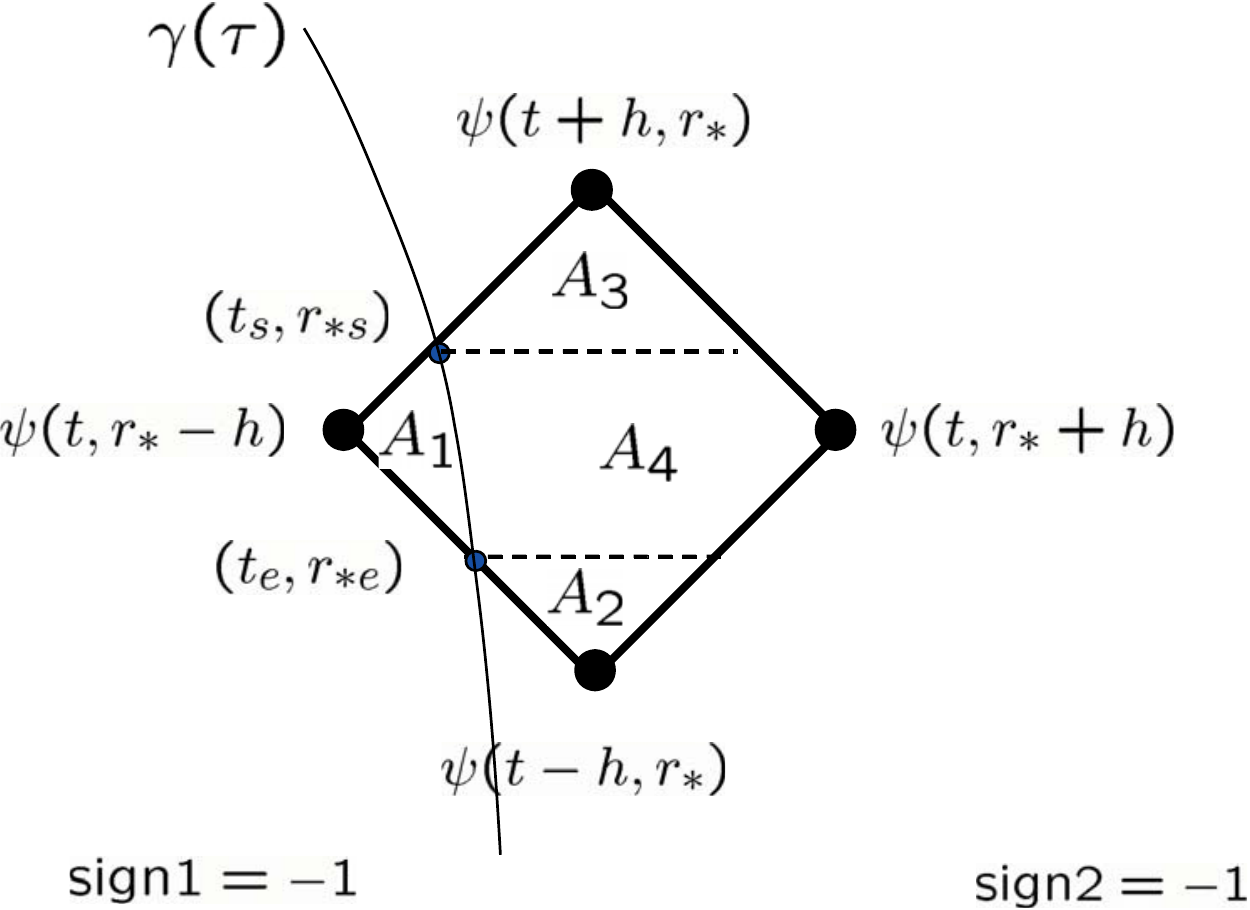}\hfill\\
\includegraphics[width=8cm]{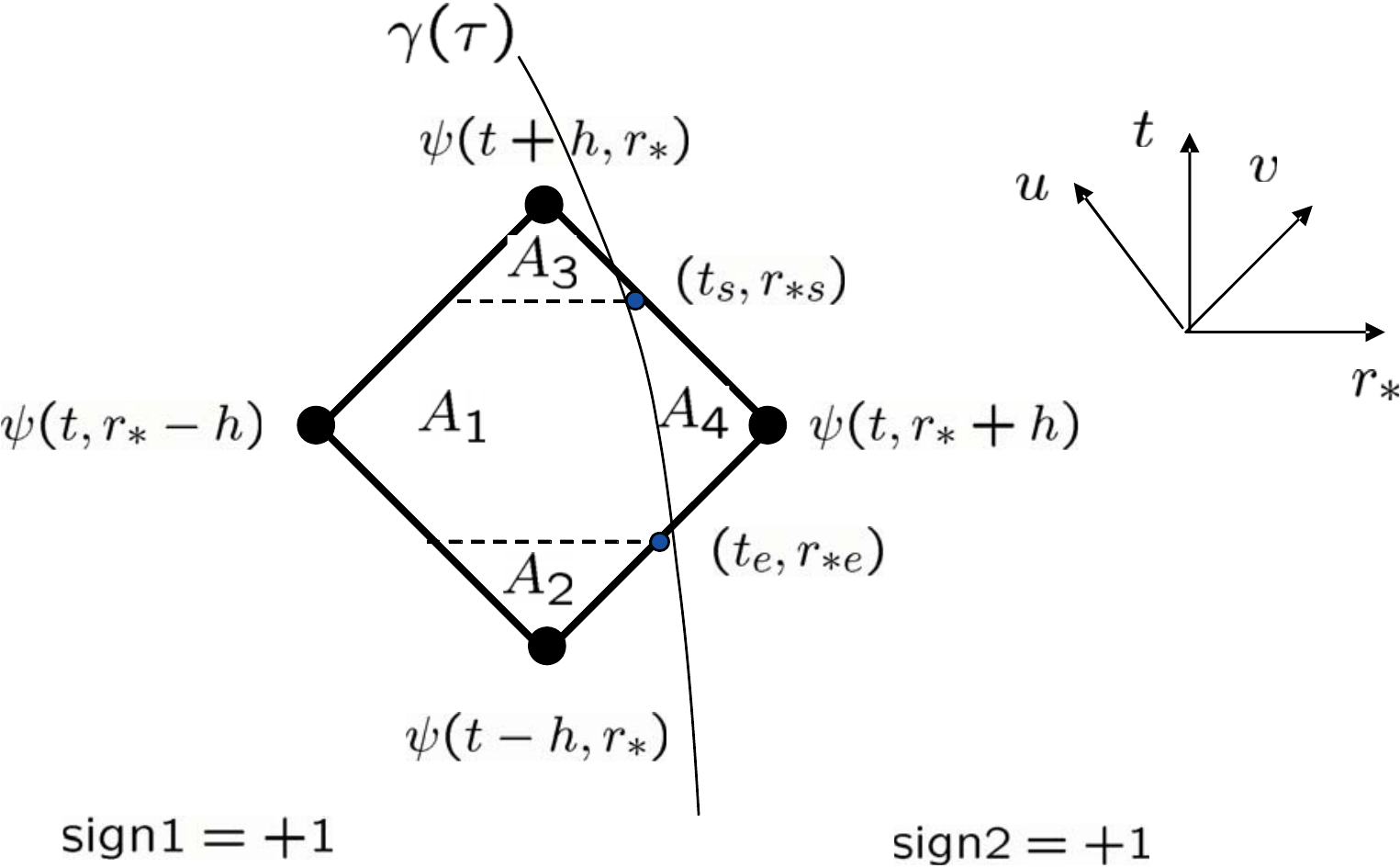}\hfill\\
\includegraphics[width=6cm]{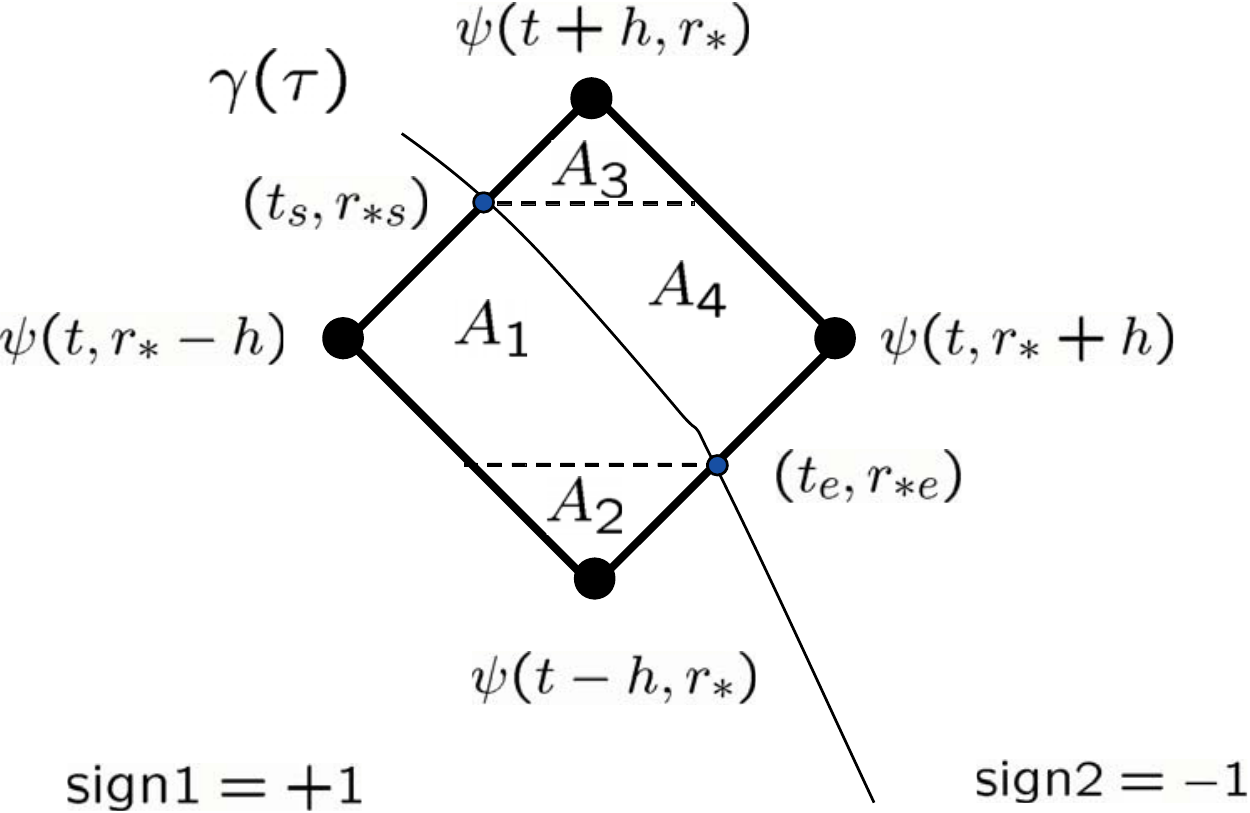}
\caption{There are only three physical cases for the particle crossing the $t,r$ cell.\label{diffcas}}
\end{center}
\end{figure}

\[
\Psi(t+h,r^*) = 
\]
\[
- \Psi(t-h,r^*) \left[
1+\frac{V(r)}{4}(A_2-A_3)\right] 
 + \Psi(t,r^*+h) \left[ 1-\frac{V(r)}{4}(A_4+A_3)\right] +
\]
\beq
\Psi(t,r^*-h,) \left[ 1-\frac{V(r)}{4}(A_1+A_3)\right] 
- \frac{1}{4} \left( 1- \frac{V(r)}{4}A_3\right) 
\int\!\!\!\!\int_{\tiny Cell} S(t,r)dA.
 \label{schemceltraversee}
\eeq

For the value of $\Psi$ at $t=h$, the unavailability of
$\Psi$ at $t=-h$, is circumvented by using a Taylor expansion of $\Psi (r^*,-h)$ for the initial conditions $t=0$:

\beq
\Psi(r^*,-h)= \Psi(r^*,0) - h \frac{\partial \psi}{\partial
 t}\mid_{t=0}.
\eeq
The setting of initial conditions constitutes a delicate, technical and largely debated issue. Apart from the technical difficulty in the numerical implementation, it suffices to state how much it is crucial to match the initial radiation conditions, that represent the earlier history of the particle, with its position and velocity. For those $z_{u0}$ starting points which are sufficiently far from the horizon, the errors on the initial conditions are fortunately not relevant at later times. 

Another numerical issue is the evaluation of the wavefunction and the perturbations at the position of the particle, but unfortunately not described in the literature and too technical for this book. The wavefunction belongs to the $C^{-1}$ continuity class\footnote{The Heaviside or step distribution, like the wavefunction of the Zerilli equation, belongs to the $C^{-1}$ continuity class; the Dirac delta distribution and its derivative belong to the $C^{-2}$ and $C^{-3}$ continuity class, respectively.} and the values before and after the particle position are computed and compared to the jump conditions
posed on the wavefunction and its derivatives \cite{ri10}. Further, it is necessary to obtain 
the third derivatives of the wavefunction to determine the correction to the geodesic background motion of the particle. Given the second order convergence of the above described algorithm, this isn't easily achieved without recurring to a fourth order scheme \cite{lo05}.
        
\section{Relativistic radial fall affected by the falling mass}

\subsection{The self-force}

It has been addressed in the previous section, that the perturbative two-body problem, involving a black hole and a particle with radiation emission, has been tackled almost 40 years ago. For computation of radiation reaction, it may be worth recalling that before 1997, only pN methods existed in the weak field regime. 
Indeed, it is only slightly more than a decade, that we possess methods \cite{misata97, quwa97} for the evaluation of the self-force\footnote{A point-like mass $m$ moves along a geodesic of the background spacetime if $m\rightarrow 0$; if not, the motion is no longer geodesic. It is sometimes stated that the interaction of the particle with its own gravitational field gives rise to the self-force. It should be added, though, that such interaction
is due to an external factor like a background curved spacetime or a force imposing an acceleration on the mass. In other words, a single and unique mass in an otherwise empty universe cannot experience any self-force. Conceptually, the self-force is thus a manifestation of non-locality 
in the sense of Mach's inertia \cite{ma83}.} in strong 
field for point particles, thanks to concurring situations. On one hand, theorists progressed in understanding radiation reaction and obtained formal prescriptions for its determination and, on the other hand, the appearance of requirements from the LISA (Laser Interferometer Space Antenna) project \cite{li} for the detection of captures of stars by supermassive black holes (EMRI, Extreme Mass Ratio Inspiral), notoriously affected by radiation reaction. 

Such factors, theoretical progress and experiment requirements, have pushed the researchers to turn their efforts in finding an efficient and clear implementation of the theorists prescriptions\footnote{It is currently believed that the core of most galaxies
host supermassive black holes on which stars and other compact objects in the neighbourhood inspiral-down and plunge in. Gravitational waves might also be detected when radiated by the Milky Way Sgr*A, the central black hole of more than 3 million solar masses \cite{fr03, cacagoko06}.
The EMRIs are further characterised by a huge number of parameters that, when spanned over a large period, produce a yet unmanageable number of templates. Thus, in alternative to matched filtering, other methods based on Covariance or on Time \& Frequency analysis are investigated. If the signal from a capture is not individually detectable, it still may contribute to the statistical background \cite{bacu04}.} by tackling the problem in the context of perturbation theory, for which the small mass $m$ corrects the geodesic equation of motion on a fixed background via a factor ${\cal O}(m)$ (for a review, see Poisson \cite{depo04} and Barack \cite{ba09}). 

Before the appearance of the self-force equation and of the regularisation methods, the main theoretical unsolved problem was represented by the infinities of the perturbations at the particle's position. 
%% For a particle represented by a Dirac delta distribution, the stress-energy momentum tensor is given by eq. (\ref{emt}).  
%% \beq
%% T_{\alpha\beta}=
%% m \int \frac{u_\alpha u_\beta~\delta[x^\gamma-z_u(\tau)]}{\sqrt{-g}}\,d\tau.
%% \eeq
%% where $g_{\alpha\beta}$, background metric, has determinant $g$; $z_u(\tau)$ is the unperturbed trajectory of the particle as %% function of proper time and $u^\alpha={dz_u}/{d\tau}$ is its four-velocity.
After determination of the perturbations through eqs. (\ref{eq:rwz*}, \ref{eq:K}-\ref{eq:H1}), the trajectory  
of the particle could be corrected simply by requiring it to be a geodesic of the total (background
plus perturbations) metric (the Christoffel connection ${\bar \Gamma}^\mu_{\alpha\beta}$ refers to the full metric):

\beq
\frac{d^2x^\alpha}{d\tau^2}+{\bar \Gamma}^\alpha_{\beta\gamma}\,\frac{dx^\beta}{d\tau}
\,\frac{dx^\gamma}{d\tau}=0,
\eeq
but the perturbation behaves as:
\beq
h_{\alpha\beta}\sim \frac{1}
{\sqrt{\left(g_{\gamma\delta}+u_{\gamma}u_{\delta}\right)
\left [x^\gamma-z_u^\gamma(\tau)\right]
\left [x^\delta-z_u^\delta(\tau)\right]
}},
\eeq
thus diverging as the inverse of the distance to the particle and imposing a singular   
behavior to ${\bar \Gamma}^\mu_{\alpha\beta}$ on the trajectory of the particle. Thus, the small perturbations assumption breaks down near the particle, exactly 
where the radiation reaction should be computed.

The solution was brought by the self-force equation, formulated in 1997 and baptised MiSaTaQuWa\footnote{In 2002 at the Capra Penn State meeting 
by Eric Poisson.}, from the surname first two initials of its discoverers, who determined it using various approaches, all yielding the same formal expression.    
In the MiSaTaQuWa
prescription, the self-force is only well defined in the harmonic (de Donder) \cite{dd21, dd27} gauge (stemmed from the Lorenz gauge \cite{lo67}) and any departure from it - its relaxation - undermines the validity of the
equation of motion.  

Mino, Sasaki and Tanaka \cite{misata97} used two methods, namely the conservation of the total stress-energy tensor and the matched asymptotic expansion. The former generalises the analysis of DeWitt and Brehme \cite{dwbr60} and Hobbs \cite{ho68}, consisting in the calculation of the electromagnetic self-force in curved spacetime previously performed in flat space by Dirac \cite{di38}. It evaluates the perturbation near the worldline using the Hadamard expansion \cite{ha23} of the 
retarded Green function \cite{gr50, gr52, gr54}. Then, it deduces the equation of motion by imposing the conservation of the rank-two symmetric total stress-energy tensor, via integration of its divergence over the interior of a thin world tube around the particle's worldline. 

The latter, reformulated by Poisson \cite{po04}, in a buffer zone matches asymptotically the expansion of the black hole perturbed background by the particle with the expansion around the particle distorted by the black hole. 

Also in 1997, the axiomatic approach by Quinn and Wald \cite{quwa97} was presented. To them, the self-force is identified by comparison of the perturbation 
in curved spacetime with the perturbation in flat spacetime. The procedure allows elimination of the divergent part and extraction of the finite part 
of the force. 

On the footsteps of Dirac's definition of radiation reaction, in 2003 Detweiler and Whiting \cite{dewh03}, see also Poisson \cite{po04}, offered a novel approach. In flat spacetime, the radiative Green function is obtained by subtracting the singular contribution, half-advanced plus half-retarded, from the retarded Green function. 
In curved spacetime, and in the gravitational case, the attainment of the radiative Green function passes through the inclusion of an additional, purposely built, function. The singular part does not exert any force on the particle, upon which only the regular field acts \cite{de01}. The latter, solely responsible of the self-force, satisfies the homogeneous wave equation and may be considered a radiative field in interaction with the particle. This approach emphasises that the motion is a geodesic of the full metric and it implies two notable features: the regularity of the radiative field and the
avoidance of any non-causal behaviour\footnote{Given the elegance of this classic approach, the self-force expression should be rebaptised as MiSaTaQuWa-DeWh.}.    

Gralla and Wald have attempted a more rigorous way of deriving a gravitational self-force \cite{grwa08}.
Their final prescription, namely self-consistency versus the first order perturbative correction to the geodesic of the background spacetime, shall be addressed later in this chapter. 
On the same track of improving rigour, an alternative approach and a new derivation of the self-force have been proposed by Gal'tsov and coworkers \cite{gaspst06} and by Pound \cite{po10}, respectively.

The determination of the self-force has allowed not only targeted applications geared to more and more complex astrophysical scenarios, but also fundamental investigations: on the role of passive, active and inertial mass by Burko \cite{bu05}; the already quoted papers on the Newtonian self-force \cite{depo04}, on the EP \cite{dewh03, whde03}, on the relation to energy conservation by Quinn and Wald \cite{quwa99}; on the relation between self-force and radiation reaction examined through gauge dependence and adiabaticity by Mino \cite{mi05a, mi05b, mi05c, mi06}; the differentiation
between adiabatic, secular and radiative approximations as well as the relevance of the conservative effects by Pound and Poisson \cite{popo08};   
on the relation between $h^{\tiny tail}$ and $h^R$, tail and regular parts of the field by Detweiler \cite{de05}.

The following wishes to be constrained to a physical and sketchy picture of the self-force. For this purpose, the original MiSaTaQuWa approach - the force acts on the background geodesic - is more intuitive. One pictorial description refers to a particle that crosses the curved spacetime and thus generates gravitational waves. These waves are partly radiated to infinity (the instantaneous part) and partly scattered back by the black hole potential (the non-local part), thus forming tails which impinge on the particle and give origin to the self-force. 
Alternatively, the same phenomenon is described by an interaction particle-black hole generating a field which behaves as outgoing radiation in the wave-zone and thereby extracts energy from the particle. In the near-zone, the field acts on the particle and determines the self-force which impedes the particle to move on the geodesic of the background metric. 
The total force is thus written as:

\beq
F^\alpha_{full}(x)= F^\alpha_{inst.}+F^\mu_{tail},
\eeq
where $F^\alpha_{inst.}$ is computed from the contributions that propagate
along the past light cone and $F^\alpha_{tail}$ has the
contributions from inside the past light cone, product of the
scattering of perturbations due to the motion of the particle in
the curved spacetime created by the black hole\footnote{Detweiler and Whiting \cite{dewh03} refer to the contribution inside the light cone via the Hadamard expression \cite{ha23} of the Green function.}, fig. \ref{sf} (pictorially, in a curved spacetime, the radiation is not solely confined to the wave front). The self-force is then computed by taking the limit
$F^\mu_{self}=F^\mu_{tail}\left[x\to z_u(t)\right]$. Thus, it is conceived as force acting on the background geodesic \cite{misata97, quwa97}, wherein $\Gamma^\alpha_{\beta\gamma}$ refers to the background metric:

\begin{figure}
\begin{center}
\includegraphics[width=2.2cm]{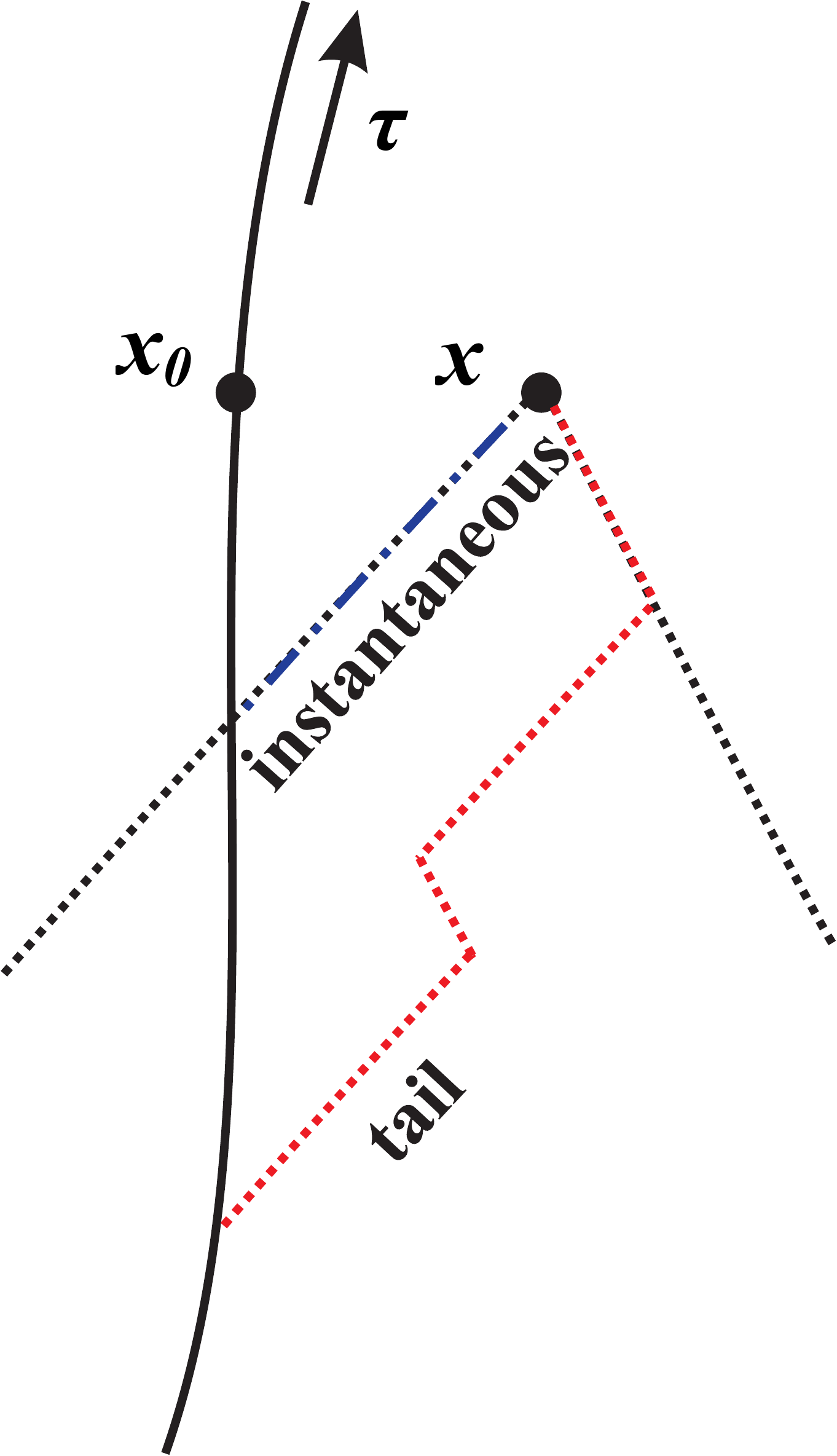}
\end{center}
\caption{The radiation is going to infinity (instantaneous) or is scattered back (tail). The latter part determines the self-force. The self-force is defined for $x$ to $ \rightarrow z_u$ where $z_u$ is the position of the particle on the worldline $\tau$, while $x$ is the evaluation point.} 
\label{sf}
\end{figure}

\beq
F^\alpha_{self} =  m
\frac{Du^\alpha}{d\tau}= \frac{d^2 x^\alpha}{d\tau ^2}+\Gamma^\alpha_{\beta
\gamma} u^\beta u^\gamma.
\label{eq:sfformal}
\eeq

All MiSaTaQuWa-DeWh approaches produce the same equation for the self-acceleration, given by: 

\beq
a^\alpha_{self} = - (g^{\alpha\beta} + u^\alpha u^\beta)\left(\nabla_\delta
h_{\beta\gamma}^* - \frac{1}{2} \nabla_\beta h_{\gamma\delta}^*\right )u^\gamma u^\delta, 
\label{eq:asf}
\eeq
where the star indicates the tail (MiSaTaQuWa) or radiative (DeWh) component. Eq. (\ref{eq:asf}) is not gauge invariant and depends upon the de Donder gauge condition:

\beq
{\tilde h}^{\mu\nu\,*}_{;\nu}=0,
\eeq
where ${\tilde h}_{\gamma\delta}^* = h_{\gamma\delta}^* - \frac{1}{2} g_{\gamma\delta}h^*$ and $h^*=g^{\mu\nu}h_{\mu\nu}^*$. Barack and Ori have shown \cite{baor01} that under 
a coordinate transformation of the form $x^\alpha \rightarrow x^\alpha - \xi^\alpha$, 
under which the perturbation transforms according to:

\beq 
h_{\mu\nu}\to h_{\mu\nu}+\xi_{\mu;\nu}+\xi_{\nu;\mu},
\eeq
the self-force acceleration transforms as:

\beq 
a^{\alpha}_{self} \to a^{\alpha}_{self}- \left[
(g^{\alpha\lambda}+u^{\alpha}u^{\lambda})\frac{d^2{\xi}_{\lambda}}{d\tau^2}
+R^{\alpha}_{\ \mu\lambda\nu}u^{\mu}\xi^{\lambda}u^{\nu}
\right],
\eeq
where the terms are evaluated at the particle and $R^{\alpha}_{\ \mu\lambda\nu}$ is the Riemann tensor
of the background geometry. Thus, for a given two-body system, the MiSaTaQuWa-DeWh acceleration is to be mentioned together with the chosen gauge\footnote{The self-force being affected by the gauge choice, the EP allows to find a gauge where the self-force disappears. Again, as in Newtonian physics, such gauge will be dependent of the mass $m$, impeding the uniqueness of acceleration.}.

The identification of the tail and instantaneous parts was not accompanied by a prescription of the cancellation of divergencies, which indeed arrived three years later thanks to the mode-sum method by Barack and coworkers \cite{baor00, ba00, ba01, babu00}. The mode-sum method relies on solutions to interwoven difficulties, mostly related to the divergent nature of the problem, but tentatively presented as separate hereafter. 

Spherical symmetry allows the force to be expanded into spherical harmonics and turns out to 
be once more the key factor for black hole physics, after having been the expedient for the determination of the wave equation. The divergent nature of the problem is then transformed into a summation problem. 
For each multipole, the full force is finite and the divergence appears only upon infinite summing over $l,m$. 

Furthermore, the tail component can't be calculated directly, but solely as difference between the full force and the instantaneous part; thus, the self-force is computed as:  

\beq
F_{self}^\alpha=\lim_{x\to z_u}\sum_{l}\left[F_{full}^{\alpha\,l\pm}(x)-F_{inst.}^{\alpha\,l\pm}(x)\right].
\label{eq:diff-for-fself}
\eeq

Each of the two quantities $F_{full}^{\alpha\,l}(x)$ and $F_{inst.}^{\alpha\,l}$ is discontinuous through the particle location and the superscript $\pm$ indicates the two (different) values obtained by taking the particle limit
from outside ($x \rightarrow z^+_u$) and inside ($x \rightarrow z^-_u$). However, the difference in eq. (\ref{eq:diff-for-fself}) does not depend upon the direction from which the limit is taken. 

The full and the instantaneous parts have the same singular behaviour at large $l$ and close to the particle; their difference should be sufficient to ensure 
a regular behaviour at each $l$. Unfortunately, another obstacle arises from the difficulty of calculating the instantaneous part mode by mode. Therefore, 
the divergence is dealt with by seeking a function $H^{\alpha\,l}$, such that the series 
$\sum_{l}\left[F_{full}^{\alpha\,l\pm}-H^{\alpha\,l\pm}\right]$ is convergent. 

The function $H^{\alpha\,l}$ mimics the instantaneous component 
at large $l$ and close to the particle. Once such condition is ensured, eq. (\ref{eq:diff-for-fself}) is rewritten 
as:

\beq
F_{self}^\alpha=\sum_{l}\left[F_{full}^{\alpha\,l\pm}-H^{\alpha\,l\pm}\right] -D^{\alpha\,l\pm},
\label{eq:sfms} 
\eeq
where
\beq
D^{\alpha\,l\pm}(x)= \lim_{x\to z_u}\sum_{l}\left[F_{inst.}^{\alpha\,l\pm}(x) - H^{\alpha\,l\pm}(x)\right].
\eeq
The addition and subtraction of the function $H^{\alpha\,l}$ guarantees the pristine value of the computation. In general, for $L = l + 1/2$:

\beq
H^{\alpha\,l} = A^\alpha\,L+B^\alpha+C^\alpha/L.
\eeq

Thus, the mode-sum amounts to \cite{ba01}: i) numerical computation of full modes; ii) derivation of the regularisation parameters $A, B, C$, and $D$ (obtained on a local analysis of the Green's function near coincidence, $x \rightarrow z_u$, at large l); iii) computation of eq. (\ref{eq:sfms}) whose behaviour has to show a $1/L^{2}$ fall off if previous steps are correctly carried out.  

\subsection{The pragmatic approach}

The straightforward pragmatic approach by Lousto, Spallicci and Aoudia \cite{lo00, lo01, sp99, spao04} is the direct implementation of the geodesic in the full metric (background + perturbations) and it is coupled to the renormalisation by the Riemann-Hurwitz $\zeta$ function. These two features justify the pragmatic adjective. Though the application of the $\zeta$ function is somewhat artificial 
and the pragmatic method is somewhat naive, the latter has the merit of: i) a clear identification of the different factors 
participating in the motion; ii) potential applicability to any gauge and to higher orders of the $\zeta$ function renormalisation. 

Dealing only with time and radial components, two geodesic equations can be written and then combined into a single one, after elimination of the geodesic
parameter. Thus, for radial fall the coordinate acceleration is given by the sole radial component:

\beq
{\ddot z}_p = 
{\bar \Gamma}_{rr}^t {\dot z}^3_p + 2{\bar \Gamma}_{tr}^t {\dot z}^2_p -{\bar \Gamma}_{rr}^r{\dot z}^2_p + {\bar \Gamma}_{tt}^t {\dot z_p} -2{\bar \Gamma}_{tr}^r{\dot z_p}- {\bar \Gamma}{tt}^r,  
\label{eq:geolou}
\eeq
where ${\bar \Gamma}^\alpha_{\beta\gamma}$ refers to the full metric and $z_p$ is given by eq. (\ref{eq:defdeltaz}). Eq. (\ref{eq:geolou}) refers to:

\begin{itemize}
\item{the full metric field ${\bar g}_{\mu\nu}(t,r)$ previously defined;} 
\item{the displacement $\Delta z$, difference between the perturbed $z_p(t)$ and the unperturbed $z_u(t)$ positions, and the coordinate time derivatives:
\beq
z_p = z_u + \Delta z,
\label{eq:defdeltaz}
~~~~~~~~~~~~~~
{\dot z}_p = {\dot z}_u + {\Delta \dot z}
\label{eq:der1deltaz},
~~~~~~~~~~~~~~
{\ddot z}_p = {\ddot z}_u + {\Delta \ddot z}~;
\label{eq:der2deltaz}
\eeq
} 
\item{the Taylor expansion of the field and its spatial derivative: 
\[
{\bar g}_{\mu\nu}\mid_{z_p} = {\bar g}_{\mu\nu}\mid_{z_u(t)} + \Delta z {\bar g}_{\mu\nu,r}\mid_{r=z_u(t)},
\]
\beq
{\bar g}_{\mu\nu,r}\mid_{z_p} = {\bar g}_{\mu\nu,r}\mid_{z_u(t)} + \Delta z {\bar g}_{\mu\nu,rr}\mid_{r=z_u(t)}.
\eeq
} 
\end{itemize}
The unperturbed trajectory of the particle $z_u(t)$ is given by the inverse of the relations $T(r)$, e.g. eq. (\ref{eq:tofr}).
Supposing that the relative strengths of the perturbations and the deviations behave as:

\beq
\frac{[h^{(1)}]^2}{g}  \simeq
\frac{h^{(2)}}{g} \ll \frac{h^{(1)}}{g} \simeq
\frac{\Delta {\dot z}}{{\dot z}_p} \simeq
\frac{\Delta  z}{z_p}.  
\label{ext.1}
\eeq
Then, the coordinate acceleration correction is given by an expansion up to $1^{st}$ order for all quantities, which corresponds to the expression in \cite{lo00,lo01}\footnote{Apart from some editorial errors therein, $\alpha_{1,2,6}$ correspond to the $A,B,C$ coefficients in \cite{lo00, lo01}, which are not to be confused with the $A,B,C$ coefficients of the mode-sum!}: 

\beq
\Delta {\ddot z} =
\alpha_1 \left(g,{\dot z}_u\right)\Delta z 
+ \alpha_2 \left(g,{\dot z}_u\right)\Delta {\dot z} 
+ \alpha_6\left(h,{\dot z}_u\right).
\label{eq:prag1}
\eeq

The particle determines in first instance the emission of radiation $h_{\alpha\beta}$, which 
after backscattering by the black hole potential, interacts 
with the particle itself resulting into a change in acceleration (the coefficient $\alpha_6$ depending on $h_{\mu\nu}$ and derivatives). The latter places the particle elsewhere from where it should have been, that is $z_u(t)$. The field is thus to be evaluated at this new position resulting into a further variation in acceleration (the terms $\alpha_1 \Delta z$ and $\alpha_2 \Delta {\dot z}$ depending on $g_{\mu\nu}$ and derivatives). 

All terms in eqs. (\ref{eq:prag1}, \ref{eq:prag2}) are of $1/M$ order; the terms $\alpha_1 \Delta z$ and $\alpha_2 \Delta {\dot z}$ represent the background field evaluated on the perturbed trajectory; $\alpha_6$ represents the perturbed field on the background trajectory. The expressions in tab.1 are gauge independent, while in tab.2 they are shown in the Regge-Wheeler gauge ($H_0=H_2$ and $K=0$ as in head-on geodesics). Finally, the coefficient $\alpha _0$ is the lowest order term corresponding to a particle radially falling into the SD black hole and not affected by the perturbations. It corresponds to the unrenormalised acceleration and it is to be added to the terms of eqs. (\ref{eq:prag1},\ref{eq:prag2}) to compute the total acceleration:

\beq
\alpha_0 = g^{tt}g_{tt,r}\dot{z}_u^2 - \frac{1}{2}g^{rr}g_{rr,r}\dot{z}_u^2 + \frac{1}{2}g^{rr}g_{tt,r} = 
\ds-\frac{M}{r^2}\left [1-\frac {2M}{r} - 3 \left(1-\frac {2M}{r}\right)^{-1}\dot{z}_u ^2\right ].
\eeq

If $\Delta {\ddot z}$ receives its main contribution from the background metric $g_{\mu\nu}$ or else cumulative effects are let to grow, a different expansion may be considered\footnote{Supposing that the relative strengths of the perturbations and the deviations behave as:
  
\beq
\frac{[h^{(1)}]^2}{g}  \simeq
\frac{h^{(2)}}{g} \ll \frac{h^{(1)}}{g} < 
\frac{\Delta {\dot z}}{{\dot z}_p} \simeq
\frac{\Delta  z}{z_p},  
\eeq
then, the coordinate acceleration correction would be given by an expansion up to 
$1^{st}$ order in perturbations and $2^{nd}$ order in deviation \cite{spao04}: 
 
\[
\Delta {\ddot z} = 
\alpha_1 \left(g,{\dot z}_u\right)\Delta z + 
\alpha_2 \left(g,{\dot z}_u\right)\Delta {\dot z} + 
\alpha_3 \left(g,{\dot z}_u\right)\Delta z^2 + 
\alpha_4 \left(g\right)\Delta {\dot z}^2 + 
\]
\beq
\alpha_5 \left(g,{\dot z}_u\right)\Delta z \Delta {\dot z} + 
\alpha_6 \left(h,{\dot z}_u\right)+ 
\alpha_7 \left(h,{\dot z}_u\right)\Delta z + 
\alpha_8 \left(h,{\dot z}_u\right)\Delta {\dot z}.
\label{eq:prag2}
\end{equation}
In eq. (\ref{eq:prag2}): i) solely second order terms in perturbations are not considered; ii) the terms 
$ \alpha_2 \left(g,{\dot z}_u\right)\Delta {\dot z}, \alpha_3 \left(g,{\dot z}_u\right)\Delta z^2, 
\alpha_4 \left(g\right)\Delta {\dot z}^2,  \alpha_5 \left(g,{\dot z}_u\right)\Delta z \Delta {\dot z}$ represent the background field evaluated on the perturbed trajectory at second order in deviation; iii) $\alpha_{3-5}$ tend to infinity close to the horizon, conversely to the $\alpha_{1,2}$ coefficients; iv) $ \alpha_7 \left(h,{\dot z}_u\right)\Delta z, \alpha_8 \left(h,{\dot z}_u\right)\Delta {\dot z}$ represent the perturbed field on the perturbed trajectory, and the $\alpha_{7-8}$ coefficients are larger near the horizon. These last two coefficients may be regularised in $l$ by the Riemann-Hurwitz $\zeta$ function as shown in \cite{spao04}.}.

\begin{table}
\caption{Representation of the terms of eq. (\ref{eq:prag1}) in gauge independent form. The elements in each column are produced by the terms of eq. (\ref{eq:geolou}), in the first row. From the algebraic sum of the elements between horizontal lines, the terms of eq. (\ref{eq:prag1}) are derived.}
\label{tab:rfgt1}
\vskip 10pt \small\rm
%\begin{tabular}{p{3cm}p{3cm}p{3cm}p{3cm}p{3cm}p{3cm}}
{\begin{tabular}{l | c | c | c | c | c | c |}  
&
$ {\bar \Gamma}_{rr}^t {\dot z}^3_p$ & 
$ 2{\bar \Gamma}_{tr}^t {\dot z}^2_p$ &
$ -{\bar \Gamma}_{rr}^r {\dot z}^2_p$ & 
$ {\bar \Gamma}_{tt}^t{\dot z}_p$ &
$ -2{\bar \Gamma}_{tr}^r{\dot z}_p$ &
$ - {\bar \Gamma}_{tt}^r$ \\ [12pt] \hline 
$ \alpha_1 \Delta z $ & &
$ g^{tt}_{,r}g_{tt,r}\dot{z}_u^2 \Delta z$& 
$ -\frac{1}{2}g^{rr}_{,r}g_{rr,r}\dot{z}_u^2 \Delta z$ & & & 
$ \frac{1}{2}g^{rr}_{,r}g_{tt,r} \Delta z$ \\ [6pt] & &
$ g^{tt}g_{tt,rr}\dot{z}_u^2 \Delta z$ & 
$ - \frac{1}{2}g^{rr}g_{rr,rr}\dot{z}_u^2 \Delta z$ & & & 
$ \frac{1}{2}g^{rr}g_{tt,rr} \Delta z$\\ [6pt] \hline 
$ \alpha_2 \Delta \dot{z}$ & & 
$ 2g^{tt}g_{tt,r}\dot{z}_u \Delta \dot{z}$ & 
$ - g^{rr}g_{rr,r}\dot{z}_u \Delta \dot{z}$ & & & \\[6pt] \hline 
$ \alpha_6$ & 
$ g^{tt}h_{tr,r}\dot{z}_u^3$& 
$- h^{tt}g_{tt,r}\dot{z}_u^2$&
$ \frac{1}{2}h^{rr}g_{rr,r}\dot{z}_u^2$&
$ \frac{1}{2}g^{tt}h_{tt,t}\dot{z}_u$& 
$ - g^{rr}h_{rr,t}\dot{z}_u$&
$ - g^{rr}h_{tr,t}$ \\[6pt] &
$ -\frac{1}{2}g^{tt}h_{rr,t}\dot{z}_u^3$&
$ g^{tt}h_{tt,r}\dot{z}_u^2$& 
$ - \frac{1}{2}g^{rr}h_{rr,r}\dot{z}_u^2$&
$ \frac{1}{2}h^{tr}g_{tt,r}\dot{z}_u$& $h^{tr}g_{tt,r}\dot{z}_u$&
$ - \frac{1}{2}h^{rr}g_{tt,r}$ \\[6pt] &
$ - \frac{1}{2}h^{tr}g_{rr,r}\dot{z}_u^3$ & & & & &
$ \frac{1}{2}g^{rr}h_{tt,r}$ \\[6pt] \hline \\
\end{tabular}}
\end{table}

\begin{table}
\caption{Representation of the terms of eq. (\ref{eq:prag1}) in Regge-Wheeler gauge. The elements in each column are produced by the terms of eq. (\ref{eq:geolou}), in the first row. From the algebraic sum of the elements between horizontal lines, the terms of eq. (\ref{eq:prag1}) are derived.}
\label{tab:rfgt2}
\vskip 10pt 
%\begin{tabular}{p}
{\begin{tabular}{l | l} 
$ \alpha _1 \Delta z$ & $\ds-\frac{M}{r^2}\left[
{ \ds\frac {6M}{r^2}} - \frac{2}{r} + {\ds\frac {6(r-M)}{r^2}\left(1-\frac {2M}{r}\right)^{-2}} \dot{z}_u ^2
\right] \Delta z$ 
\\[12pt] \hline \\ 
$\alpha _2 \Delta \dot{z}$ &  ${ \ds\frac {6M}{r^2}\left(1-\frac {2M}{r}\right)^{-1}}\dot{z}_u \Delta \dot{z}$
\\[12pt] \hline \\
$\alpha _6$ &  
$ \ds\frac{1}{r - 2M}
\left[
\frac {r^2 H_{0,t}}{2(r - 2M)} - \ds\frac {MH_1}{r - 2M} - rH_{1,r} 
\right]
\dot{z}_u^3
- \ds\frac {3}{2} H_{0,r}
\dot{z}_u^2
- 3 \left(
{ \ds\frac {H_{0,t}}{2}}  - { \ds\frac {MH_1}{r^2}}
\right )
\dot{z}_u
$ 
\\[12pt] \\
& $
+ { \ds\frac {r - 2M} {r}} 
\left [ {\ds\frac{2MH_0}{r^2}} + 
{\ds\frac{(r-2M)H_{0,r}}{2r}} - H_{1,t} 
\right]
$
\\[12pt] \hline \\
\end{tabular}}
\end{table}

In radial fall, it has been indicated by two different heuristic arguments \cite{lo00, lona09} that the metric perturbations should be of $C^0$ continuity class at the location of
the particle\footnote{The jump conditions were also dealt with by Sopuerta and Laguna \cite{sola06}.}. One argument \cite{lo00} is based on the integration over $r$ of the Hamiltonian constraint, which is the $tt$ component of the 
Einstein equations (eq.[C7a] in \cite{ze70c}); the other \cite{lona09} 
on the structure of selected even perturbations equations. In \cite{aosp10a}, a stringent analysis on the $C^0$ continuity is pursued in terms of the 
jump conditions that the wavefunctions and derivatives have to satisfy to guarantee the continuity of the perturbations\footnote{ 
Having suppressed the $l$ index for clarity of notation, after visual inspection of eq. (\ref{eq:rwz*}), containing a derivative of the Dirac delta distribution, it is evinced that the wavefunction $\Psi$ is of $C^{-1}$ continuity class and thus can be written as:

\beq 
\Psi(t,r)=\Psi^+(t,r)~\Theta_1+\Psi^-(t,r)~\Theta_2,
\label{eq.conti.psienf.plus.moins}
\eeq
where $\Theta_1 = \Theta\left[r-z_u(t)\right ]$, and $\Theta_2 = \Theta\left[z_u(t) - r \right ]$ are two Heaviside step distributions. 
Computing the first and second, space and time and mixed derivatives, Dirac delta distributions and derivatives are obtained of the type  
$\delta[r-z_u(t)]$ and $\delta'[r-z_u(t)]$, respectively.
It is wished that the 
discontinuities of $\Psi$ and its derivatives are such that they are canceled when combined in $K$, $H_2$ and $H_1$. 
After replacing $\Psi$ and its derivatives in eqs. (\ref{eq:K}-\ref{eq:H1}), continuity requires that the coefficients of $\Theta_1$ must be equal to the coefficients of $\Theta_2$, while 
the coefficients of $\delta$ and $\delta '$ must vanish separately. After some tedious computing and making use  
of one of the 
Dirac delta distribution properties: $f(r)
\delta'[r-z_u(t)] = f[z_u(t)] \delta'[r-z_u(t)] - f'(z_u(t)) \delta[r-z_u(t)]$, at the position of the particle, 
the jump conditions for $\Psi$ and its derivatives are found. Furthermore, the jump conditions allow a new method of integration, as shown by Aoudia and Spallicci \cite{aosp10a}. }.
Anyhow, the connection coefficients and the metric perturbation derivatives have a finite jump and they can be computed as the average of their values at
$z_u\pm\epsilon$ with $\epsilon\to0$. 

The supposed $C^0$ continuity class of the metric perturbations allows to deal with the divergence with $l$ of the $\alpha_6$ coefficient \cite{lo00, lo01}. The divergence originates from the infinite sum over the finite multipole component contributions. One way of regularising this
sum is to subtract to each mode precisely the $l\to\infty$ contribution, since for ever larger $l$ the metric perturbations tend to some finite asymptotic behaviour.  Thus, the subtraction from each mode of the $l\rightarrow \infty$ part leads to a convergent series. The renormalisation by the Riemann-Hurwitz $\zeta$ function was proposed first in \cite{lo00, lo01} and then extended 
to higher orders in \cite{spao04}. For $L = l + 1/2$, it can be shown that:

\begin{equation}
\alpha_6 = \sum_{l=0}^\infty \alpha_6^l,
~~~~~~~~~~~~~
\alpha_6^l = \alpha_{6\pm}^a L + \alpha_6^b L^0 + \alpha^c_{6\pm} L^{-1} + \alpha_6^d L^{-2} + {\cal O}(L^{-3}).
\label{eq:cdanorm}
\end{equation}
Eq. (\ref{eq:cdanorm}) is casted to have a similar form to the mode-sum expression. The average of $\alpha^a_{6\pm}$ and $\alpha^c_{6\pm}$ vanish at the position of the particle, whereas $\sum_{l=0}^\infty \alpha_6^b = \infty$ determines the divergence.  

The Riemann $\zeta$ function \cite{ri59} and its generalisation, the Hurwitz $\zeta$ function \cite{hu82}, are defined by:

\begin{equation}
\zeta (s) = \sum_{l=1}^\infty (l)^{-s},
~~~~~~~~~~~~~
\zeta (s,a) = \sum_{l=0}^\infty (l + a)^{-s},
\end{equation}
where in our case $a = 1/2$. Two special values of the Hurwitz functions, namely $\zeta (0, 1/2) = 0$ and $\zeta (2, 1/2) = 1/2~\pi^2$, 
cancel the divergent term and determine that the term $\sum_{l=0}^\infty \alpha_6^d L^{-2}$ gets a finite value, respectively.
Barack and Lousto \cite{balo02} have shown the concordance of the mode-sum and the $\zeta$ regularisations for radial fall. 

\section{The state of the art}

It is now time to discuss the state of the art of the radial fall affected by its mass and the emitted radiation.  
As shown in the introduction, the adiabatic approximation requires that a given orbital parameter $q$ changes slowly over time scales comparable to the orbital period $P$ (this is somewhat a coarse definition since the small mass always `reacts' immediately): $\Delta q = {\dot q}P \ll q$. For circular and moderately elliptic orbits, the above condition, where $q$ is function of the semi-lactus rectum $p$ and eccentricity $e$, is transformed into a condition on the $m/M$ ratio \cite{cukepo94}. In radial fall, though, it is far from being evident, and even possible, to identify a condition on adiabaticity within which 
any simplification may occur. The feebleness of cumulative effects for radiation reaction does not imply their non-existence. On the contrary, this is the case where most care and sophisticated techniques are demanded for the computation of the motion affected by the back-action, even if the latter has moderate effects.  
Therefore it is not surprising, thanks to the feebleness and to the difficulties, that solely two studies (one based on the pragmatic method, the other on the self-force) exist.

\subsection{Trajectory}

The perennial question on the behaviour of the infalling mass reflects itself in the determination along which direction the back-action is exerted.   

Lousto (fig. 2a in \cite{lo00}) suggests that the $\alpha_6$ term, denoted therein $C$ (the variation of the coordinate acceleration of the particle due solely to perturbations; it  corresponds to the self-force when referred to coordinate time, see next section), increases approaching the Zerilli potential at $3.1 M$ and reaches its peak value around $2.4 M$. The same reference (fig. 2b in \cite{lo00})  
shows that the coordinate acceleration, thus including  $\alpha_1 \Delta z$ and $\alpha_2 \Delta {\dot z}$ terms, is slowed down\footnote{Lousto \cite{lo00} comments only this former part and not the acceleration boost taking place after the Zerilli potential peak.} and mostly until around $3.1 M$. The two statements are not contradictory if the discrepancy is attributed to $\alpha_1 \Delta z$ and $\alpha_2 \Delta {\dot z}$. In the same reference \cite{lo00}, deceleration is expected as the system is losing energy and momentum. 
This repulsive behaviour - before the Zerilli potential peak - is confirmed in the abstract of \cite{lo01}. 

Conversely, for Barack and Lousto \cite{balo02} the radial component of the self-force is found to point inward (i.e., toward the black hole) throughout the entire plunge, independently on the starting point 
$z_{u0}$. The work done by the self-force is considered positive, resulting in an increase of the energy parameter $E$ throughout
the plunge. To these results, it is attached a specific gauge choice (as opposed to the energy flux at infinity, which is gauge invariant) \cite{balo02}.

The upward or inward direction impressed on the particle by its mass and the emitted radiation and whether this direction is maintained throughout the plunge or part of it, is a fundamental if not the main one, feature to acquire in such analysis. Again here, the statements from the three papers might not be contradictory as Lousto
\cite{lo00,lo01} describes motion in coordinate Schwarzschild time, and includes geodesic deviations. Conversely, Barack and Lousto \cite{balo02} describe motion in proper time and apply only the self-force without geodesic deviations. 
Nevertheless, it is of interest to remark that the concept of repulsion resurfaces again solely in coordinate time as in the elder debate. The coefficient of the geodesic deviation coefficient $\alpha_1$ changes sign during fall, while it does not occur to $\alpha_2$ \cite{aosp10b} which remains negative throughout.
    
For a particle
starting from rest at a finite distance from the black hole, an analytic approximation of the self-force for the $l\geq 2$ modes (while for $l=0,1$ the solutions in \cite{ze70c} are mentioned) is given by Barack and Lousto\cite{balo02}: 

\beq
F_{self}^r=F_{self}^{r\,l=0}+F_{self}^{r\,l=1}+
\sum_{l=2}^{\infty} \left\{-\frac{15}{16}m^2\frac{E^2}{r^2}
\left(E^2+\frac{4M}{r}-1\right)\frac{1}{L^2}+{\cal O}(L^{-4})\right\},
\label{eq:sfrf}
\eeq
where $E$ is the orbital energy of the particle. The force has only negative and even powers of $l$,
which makes the sum quickly convergent and provides an excellent
approximation for the numerical evaluation of the first few lower multipoles. The derivation of such net expression for the self-force is not described in \cite{balo02}\footnote{In the Rapid Communication \cite{balo02} there are $7$ citations of a yet unpublished material containing mathematical and numerical justifications of the results therein. The author acknowledges private communications by L. Barack.}.

\subsection{Regularisation parameters }

Regularisation parameters of the mode-sum method have been confirmed independently by different papers \cite{ba01, baminaorsa02} and they are consistent with the results obtained by the application of the $\zeta$ function \cite{balo02}. In radial fall, there is a regular gauge transformation between the de Donder and Regge-Wheeler gauges \cite{baor01}, and thus the regularisation parameters 
were also determined in the latter gauge \cite{ao08, balo02}. The results are:

\beq
\label{A}
A_{\pm}^r=\mp \frac{m^2}{z_u^2}\,E,
~~~~~~~~~~~~
A_{\pm}^t=\mp \frac{m^2}{z_u^2}\,\frac{\dot{r}_p}{f},
~~~~~~~~~~~~
A_{\pm}^\theta=A_{\pm}^\varphi=0, 
\eeq

\beq
\label{B}
B^r=-\frac{m^2}{2z_u^2}\,E^2,
~~~~~~~~~~~
B^t=-\frac{m^2}{2z_u^2}\,\frac{E\dot{r}_p}f,
~~~~~~~~~~~~
B^\theta=B^\varphi=0,
\eeq

\beq
\label{CD}
C^{\alpha}=D^{\alpha}=0,
\eeq
where $f\equiv 1-2M/z_u$ and ${\dot z}_u =-(E^2-f)^{1/2}$. 

\subsection{Effect of radiation reaction on the waveforms during plunge}

Thanks to a suggestion of B. Whiting, preliminary indications were found \cite{ao08}. The waveforms shifts are of the order of (tens of) seconds for a particle sensibly radiating for few thousands of seconds, having started at rest from a finite distance between $4M$ and $40M$. The assumption used therein is energy-momentum balance (the energy radiated to infinity and absorbed by the black hole is imposed to be equal to the energy change in the particle fall).
This assumption is likely jeopardised by the lack of instantaneous energy conservation \cite{quwa99}. 

The correct alternative is the computation of the self-force and its continuous implementation all along the trajectory. 
It is thus mandatory to consider the application of the recently proposed self-consistent prescription \cite{grwa08}, unfortunately not yet part of the state of the art in terms of its application.  

\section{Beyond the state of the art: the self-consistent prescription} 

In \cite{grwa08} a rigorous derivation and application of the self-force equation is proposed. In the derivation, emphasis is put upon the de Donder gauge and the consequences of relaxation, that is not enforcing this gauge.
On one hand, the de Donder gauge is imposed by the nature of the self-force, solely defined in this gauge. On the other hand, the relaxation of the de Donder gauge stems from the need of departing from the background geodesic to find the self-force that causes such departure.   
Previous derivations were based on the assumption of deviations from geodesic motion expected to be small; by consequence, the de Donder gauge
violation should likewise be small. Instead, the new derivation by Gralla and Wald is a rigorous, perturbative, result, obtained without the step of de Donder gauge relaxation and containing the geodesic deviation terms. 

But Gralla and Wald go a step further \cite{grwa08}. For the evolution of an orbit, rather than a first order perturbation equation containing geodesic deviation terms, they recommend a self-consistent approach. Such prescription basically affirms the greater accuracy of a first order perturbation expansion along a continuously corrected trajectory as opposed to a higher order perturbation expansion made on the background geodesic\footnote{The evolution of an orbit is lately getting the necessary concern. Pound and Poisson \cite{popo08} apply osculating orbits to EMRI, but unfortunately their method is not applicable to plunge, for two reasons: the semi-latus rectum of the orbit, which decreases for radiation reaction, is smaller than a given quantity, considered as limit in their study case; the velocities and fields in the plunge are
highly relativistic and their post-Newtonian expansion of the perturbing force becomes inaccurate. Non-applicability to plunge stands also for the work by Hinderer and Flanagan \cite{hifl08}.}. 
Self-consistency bypasses the issue of relaxation, since at each integration step a new geodesic is found\footnote{Indeed,  
it affirms that it is preferable to apply successively a $1^{st}$ order expansion at $x_0$ and then at 
$x_1$, $x_2$, ... $x_m$, rather then a $2^{nd}$ or higher order expansion at solely $x_0$.   
It is evident, though, that 
self-consistency and perturbation order are decoupled concepts and that the former may be conceptually applicable to higher orders and more specifically, when, and if, a second order formalism will be available: in the same line of reasoning it would be preferable to apply successively a $2^{nd}$ order expansion at $x_0$ and then at $x_1$, $x_2$, ... $x_m$, rather then a $3^{rd}$ or higher order expansion at solely $x_0$.}. 

The `classic' first order perturbative expansion for the motion of a small body determines that the first order metric perturbations satisfy:

\beq
\nabla^\gamma \nabla_\gamma \tilde{h}_{\alpha\beta} - 2 R^\gamma{}_{\alpha\beta}{}^\delta \tilde{h}_{\gamma\delta} =
- 16 \pi M u_\alpha u_\beta \frac{\delta^{(3)}(x^\mu)}{\sqrt{-g}} \frac{d\tau}{dt},
\label{eq:p11}
\eeq
where $x^\mu = 0$ corresponds to a geodesic $\gamma$ of the background
spacetime, and $u^\alpha$ is the tangent to $\gamma$. It is reminded that $\tilde{h}_{\alpha\beta} \equiv h_{\alpha\beta} - \frac{1}{2} h g_{\alpha\beta}$ with $h = h_{\alpha\beta} g^{\alpha\beta}$. For a retarded solution to this equation (thus satisfying the
de Donder gauge condition) of the type:

\beq
h_{\alpha\beta}^{\tiny \textrm{tail}}(x) = M
\int_{-\infty}^{\tiny \tau_{\textrm{ret}}^-}\left(G^+_{\alpha\beta \alpha '\beta '}-\frac{1}{2}g_{\alpha\beta}G^{+ \gamma}_{\gamma\alpha '\beta '}\right) \left[x,z_u(\tau')\right ] u^{\alpha '}u^{\beta '} d\tau', 
\label{eq:p12}
\eeq
the first order in $\lambda$ deviation of the motion from $\gamma$
is expressed by (see \cite{grwa08} for the additional spin term):

\beq
u^\gamma\nabla_\gamma(u^\beta\nabla_\beta Z^\alpha) =  
- {R_{\beta\gamma\delta}}^\alpha u^\beta Z^\gamma u^\delta
- (g^{\alpha\beta} + u^\alpha u^\beta)(\nabla_\delta
h_{\beta\gamma}^{\tiny \textrm{tail}}- \frac{1}{2} \nabla_\beta h_{\gamma\delta}^{\tiny
\textrm{tail}})u^\gamma u^\delta. 
\label{eq:p13}
\eeq

Self-consistency prescribes that rather than using eqs. (\ref{eq:p11},\ref{eq:p12},\ref{eq:p13}), it is instead preferable to apply the self-force coherently all along the trajectory:

\beq
\nabla^\gamma \nabla_\gamma \tilde{h}_{\alpha\beta} - 2 R^\gamma{}_{\alpha\beta}{}^\delta \tilde{h}_{\gamma\delta} =
- 16 \pi M u_\alpha(t) u_\beta(t) \frac{\delta^{(3)}\left[x^\mu - z^\mu_p(t)\right ]}{\sqrt{-g}} \frac{d\tau}{dt},
\label{eq:sc1}
\eeq

\beq
u^\beta \nabla_\beta u^\alpha = - (g^{\alpha\beta} + u^\alpha u^\beta)(\nabla_\delta
h_{\beta\gamma}^{\tiny \textrm{tail}}- \frac{1}{2} \nabla_\beta h_{\gamma\delta}^{\tiny
\textrm{tail}})u^\gamma u^\delta, 
\label{eq:sc2}
\eeq

\beq
h_{\alpha\beta}^{\tiny \textrm{tail}}(x) = M
\int_{-\infty}^{\tiny\tau_{\textrm{ret}}^-}\left(G^+_{\alpha\beta \alpha '\beta '}-\frac{1}{2}g_{\alpha\beta}G^{+ \gamma}_{\gamma\alpha '\beta '}\right) \left[ x,z_p(\tau')\right ] u^{\alpha '}u^{\beta '} d\tau', 
\label{eq:sc3}
\eeq
where this time 
$u^\alpha(\tau)$ in eqs. (\ref{eq:sc2},\ref{eq:sc3}), normalised in the background metric, refers to the self-consistent motion $z_p(\tau)$, rather than to a background geodesic as in eqs. (\ref{eq:p12},\ref{eq:p13}); $G^+_{\alpha \beta \alpha '\beta '}$ is the retarded Green function, normalised with a factor of $-16 \pi$ \cite{quwa97}; the symbol $\tau_{\tiny\textrm{ret}}^-$ indicates the range of the integral being extended just short of the retarded time $\tau_{\tiny\textrm{ret}}$, so that only the interior part of the light-cone is used.  

The geodesic deviations vanish in eq. (\ref{eq:sc2}), since self-consistency is imposed. Nevertheless, there might be situations where, for a whatever reason, the numerical implementation of the self-consistent prescription may be cumbersome. In this case, the addition of geodesic deviation terms might as in eq. (\ref{eq:p13}) be necessary.  

If furthermore, motion is wished to be expressed in coordinate time, like in the pragmatic approach, it is useful to find a correspondence between the first order perturbation equation, containing geodesic deviations, eq. (\ref{eq:p13}), and the terms of eq. (\ref{eq:prag1}). It is going to be shown that the $\alpha_6$ term of eqs. (\ref{eq:prag1}, \ref{eq:prag2}) corresponds to the self-force term of eq. (\ref{eq:p13}), when the latter is transferred to coordinate time. 
Referring to the time and radial components of the self-force, it is obtained from eqs. (\ref{eq:sfformal}, \ref{eq:asf}): 

\beq
\frac{d^2 t}{d\tau ^2} = F^t_{self} - m \Gamma^t_{\beta\gamma} u^\beta u^\gamma =
- m g^{t\beta} \left(h^*_{\beta\gamma\,;\delta} - \frac{1}{2} h^*_{\gamma\delta\,;\beta} \right) u^\gamma u^\delta 
- m k^t, 
\label{eq:d2td2tau}
\eeq
 
\beq
\frac{d^2 r}{d\tau ^2} = F^r_{self} = - m \Gamma^r_{\beta\gamma} u^\beta u^\gamma 
- m g^{r\beta} \left(h^*_{\beta\gamma\,;\delta} - \frac{1}{2} h^*_{\gamma\delta\,;\beta} \right) u^\gamma u^\delta 
- m k^r, 
\label{eq:d2red2tau}
\eeq
where 

\beq
k^\alpha 
= \left(h^*_{\beta\gamma\,;\delta} - \frac{1}{2} h^*_{\delta \gamma\,;\beta} \right) 
u^\alpha u^\beta u^\gamma u^\delta.
\eeq
Since:

\[
\frac{d}{d\tau} = \frac{dt}{d\tau} \frac{d}{dt}~~~~~~~~~~~
\frac{d^2 r }{d\tau^2}= \frac{dr}{dt}\frac{d^2 t }{d\tau^2}+ \frac{d^2 r }{dt^2}\left (\frac{dt}{d\tau}\right )^2,
\]
after some computation, the term $k^\alpha$ disappears when the self-force is expressed in coordinate time:

\[
m \frac{d^2 r }{dt^2} = m \left[
\Gamma^t_{\beta\gamma} v^\beta v^\gamma + 
g^{t\beta} \left(h^*_{\beta\gamma\,;\delta} - \frac{1}{2} h^*_{\gamma\delta\,;\beta} \right)v^\gamma v^\delta \right]{\dot z_u(t)}
\]
\beq
- m\Gamma^r_{\beta\gamma} v^\beta v^\gamma  
- m g^{r\beta} \left(h^*_{\beta\gamma\,;\delta} - \frac{1}{2} h^*_{\gamma\delta\,;\beta} \right)v^\gamma v^\delta. 
\label{eq:d2red2t}
\eeq 
Furthermore, a tedious computation shows that eq. (\ref{eq:d2red2t}) is nothing else than the $\alpha_6$ term of eq. (\ref{eq:prag1}) apart from the regularisations by mode-sum or Riemann-Hurwitz $\zeta$ function:

\beq
\alpha_6 \leftrightarrow  
g^{t\beta} 
\left(h^*_{\beta\gamma\,;\delta} - \frac{1}{2} h^*_{\gamma\delta\,;\beta} \right)
v^\gamma v^\delta {\dot z_{u(t)}}
- g^{r\beta} 
\left(h^*_{\beta\gamma\,;\delta} - \frac{1}{2} h^*_{\gamma\delta\,;\beta} \right)
v^\gamma v^\delta. 
\eeq 

The recasting of the Riemann tensor term and of the left-hand side of eq. (\ref{eq:p13}) into coordinate time determines the relation to the $\alpha_1  \Delta z$ and $\alpha_2 \Delta {\dot z}$ terms in eqs. (\ref{eq:prag1}, \ref{eq:prag2}). Finally, the geodesic deviation equation\footnote{See Levi-Civita \cite{lc25,lc26}, Ciufolini and Wheeler \cite{ciwh95}, Ciufolini \cite{ci86}.}  eq. (\ref{eq:p13}), is dealt with by Aoudia and Spallicci \cite{aosp10b} as difference between background and perturbed motions.   

\section{Conclusions}

It has been shown that free fall is still the arena for a deeper comprehension of gravitation. Furthermore, it still generates acute observations like relativistic gliding for which the asymmetric oscillations of a quasirigid body slow down or accelerate its fall in a gravitational background \cite{gumo07}.

But is the problem of radial fall solved? Do we know the laws of motion of a star falling into a black hole, the relativistic modern version of the falling stone? A fair and objective answer leads to a moderate optimism.  
The general relativistic problem has had undeniable progress from 1997, but a careful analysis of the literature shows that some issues are either still partly open or simply not fully at hand, in terms of clear procedures, by means of which clearly cut answers are obtained. 

The remaining steps to be fulfilled for a satisfactory level of comprehension for radial fall of a small particle into a large mass (represented by a SD black hole) within the first perturbative order in $m/M$ are divided in three groups.

Investigations to be pursued before recurring to the self-consistent prescription: 

\begin{enumerate}[I]
\item{
Compute in proper time the first order perturbation equation with deviation terms, i.e. all terms of eq. (\ref{eq:p13}).}
\item{ 
Identify and compare expressions for renormalised, semi-renormalised and unrenormalised accelerations in the SD perturbed geometry; evaluate repulsive effects.
The last century debate on repulsion has not addressed the effects of mass finitude ($l=0,1$) and radiation reaction on the motion ($l\geq 2$). The inclusion of perturbations blurs the issue further and renders it more complex. 
}
\item{Compare the waveform corrections obtained with energy-balance by Aoudia \cite{ao08} with those to be obtained by the integration in the equation of motion of the approximate analytic expression of the self-force of Barack and Lousto \cite{balo02}. The comparison is of interest only in case of concurring outcomes, as both procedures have a large degree of approximation. An other comparison \cite{sp09} could be made with the numerical results for a 10 : 1 mass ratio, as larger mass ratios imply the solution of the numerical two-scale problem, not yet at hand.}
\item{Solve the initial conditions for particles with non-null or even relativistic initial velocities with and without large eccentricities. If the outcome of the previous item is successful, coarsely identify possible signatures of radiation reaction\footnote{For Mino and Brink \cite{mibr08}, the energy and momentum radiated are computed on the assumption that the small body falls in a dynamical time scale, with respect to proper time, well short of the radiation reaction time scale and therefore the gravitational radiation back-action on the orbit is considered negligible. The particle plunges on a geodesic trajectory, incidentally starting from a circular orbit, thus at zero initial radial velocity.}. It is likely that a non-null initial velocity would act as an amplifier of the waveform shifts due to radiation reaction. In \cite{lopr97a}, Lousto and Price show that the energy radiated by a non-null initial velocity may rise up to almost two orders of magnitude (see also fig. 3 in \cite{lo97}). Incidentally, this line of work is confluent with the interests shown recently in the particle physics community \cite{ardodsto08, spcaprbego08, shokya08, ve10, yosh09} for head-on collisions and the associated radiation reaction \cite{gakospto10a, gakospto10b}. Hopefully, progress in numerical relativity may lead to analysis of larger mass ratios and thereby test future results of perturbation theory\footnote{For head-on collisions, Price and Pullin have surprisingly shown the applicability of perturbation theory for the computation of the radiated energy and of the waveform for two equal black holes, starting at very small separation distances \cite{prpu94}, the so called close-limit approximation.} in the range beyond 
$10^{-3}$.}

\vskip 5pt

Investigations to be performed solely by the implementation of the self-consistent prescription:

\vskip 5pt

\item{Evaluate the trajectory by means of the Gralla and Wald self-consistent method \cite{grwa08}. At each integration step, for a given number of modes: evaluation of the perturbation functions at the position of the particle; regularisation by mode-sum or $\zeta$ methods; correction of the geodesic and identification 
of the cell crossed by the particle; computation of the source term; reiteration of the above. Perturbations and self-force analysis in de Donder's gauge are on-going \cite{balo05, basa10}.}
\item{Repeat steps 1-4 on the basis of the acquired self-consistent trajectory. An iterative scheme may be envisaged also in coordinate time.}

\vskip 5pt

Investigations to be performed independently from the self-consistent prescription: 

\vskip 5pt

\item{Identify whether in a thought radial fall experiment, there is a physical observable independent of
the gauge choice or else manifesting a recognisable effects in a given gauge, following Detweiler \cite{de08}.}
\item{Identify the domain of applicability of the self-force.  Although self-consistency represents a closer description to perturbed motion, it is limited to cases where local deviations are small, since after all, it
remains a perturbative approach. Thus, a quantitative identification of the domain of applicability of such prescription would be of interest to the community developing LISA templates. The adiabatic approximation can't be evoked to establish the limits of any self-force based analysis. Indeed, the hypothesis on the feeble magnitude of local deviations may be less constraining than the adiabatic hypothesis, for the absence of requirements on averaging. 
An explicit condition, referring to the geometry of the orbit and to the mass
ratio and defining the domain of applicability of the self-force versus fully numerical
approaches, is not yet available.} 
\item{Investigate other gauges than Regge-Wheeler and other methods than the self-force for radial fall. A confirmation of the results by Barack, Lousto in \cite{lo01, balo02} by an independent group is missing, though in \cite{ao08} the necessary basic tools are developed, e.g. a numerical programme confirming the waveforms of \cite{lopr97b, mapo02}. It would be also beneficial to carry out an analysis by the EOB method \cite{da09b}
\footnote{A non-recent analysis by Simone, Poisson and Will \cite{sipowi95}, between pN and perturbation methods in head-on collisions, was limited to the computation of the gravitational wave energy flux.}. 
}
\end{enumerate}

The self-force community and even more the EMRI community are animated by different interests ranging from fundamental physics and theory, to numerical applications, data analysis and astrophysics. 
To these variegated communities and generally to physicists and astrophysicists, the capture of stars by supermassive black holes will bring a
development comparable to the ancient markings made by the leaning Pisa tower, the Cambridge apple tree and the Einstein lift. 

\begin{acknowledgement}
Discussions throughout the years with S. Aoudia, L. Barack, S. Chandrasekhar, S. Detweiler, C. Lousto, J. Martin-Garcia, S. Gralla, E. Poisson, R. Price, R. Wald, B. Whiting are acknowledged. I would like to thank E. Lallier Verg\`es for the support to the CNRS School on the Mass and the $11^{th}$ Capra conference, held in June 2008 in Orl\'eans. It was my sincere hope that both events could put together different communities working on mass and motion: part of the contributors to this book and their colleagues (Blanchet, Detweiler, Le Tiec and Whiting) have already made concrete steps towards such cooperation \cite{bldeltwh10a, bldeltwh10b}.
\end{acknowledgement}

\end{document}